\documentclass[journal]{IEEEtran}  

\IEEEoverridecommandlockouts                              

\usepackage{mathtools}    
\usepackage{amsfonts}
\usepackage{graphicx}   
\usepackage{subcaption}
\usepackage{epsfig} 
\usepackage{cancel}
\usepackage{amssymb}
\usepackage{color}
\usepackage{bm}
\usepackage{todonotes} 
\usepackage{float}
\usepackage{cite}
\usepackage{enumitem}
\usepackage{subcaption}
\usepackage{siunitx}
\usepackage[ruled,vlined,titlenotnumbered,linesnumbered]{algorithm2e} 
\usepackage{array}

\newcommand{\R}{\mathbb{R}} 
\newcommand{\rc}{R_c} 

\newcommand{\N}{N} 

\newcommand{\veh}{Q} 
\newcommand{\intr}{I} 
\newcommand{\state}{x} 
\newcommand{\ctrl}{u} 
\newcommand{\dstb}{d} 
\newcommand{\pos}{p} 
\newcommand{\npos}{h} 

\newcommand{\fdyn}{f} 
\newcommand{\cset}{\mathcal{U}} 
\newcommand{\cfset}{\mathbb{U}} 
\newcommand{\dset}{\mathcal{D}} 
\newcommand{\dfset}{\mathbb{D}} 
\newcommand{\obsset}{\mathcal{G}} 
\newcommand{\dz}{\mathcal{Z}} 
\newcommand{\sep}{\mathcal{S}} 
\newcommand{\buff}{\mathcal{B}} 

\newcommand{\valfunc}{V} 
\newcommand{\valfuncfwd}{W} 
\newcommand{\brs}{\mathcal{V}} 
\newcommand{\frs}{\mathcal{W}} 
\newcommand{\pfrs}{\mathcal{P}} 
\newcommand{\targetset}{\mathcal{L}} 
\newcommand{\ham}{H} 
\newcommand{\fc}{l} 
\newcommand{\ic}{l} 
\newcommand{\obsfunc}{g} 
\newcommand{\costate}{\lambda}

\newcommand{\disckernel}{\Omega} 

\newcommand{\ldt}{t^\text{LDT}} 
\newcommand{\sta}{t^\text{STA}} 
\newcommand{\ioset}{\mathcal{O}} 
\newcommand{\boset}{\mathcal{M}} 
\newcommand{\soset}{\ioset^\text{static}} 
\newcommand{\iat}{t^\text{IAT}} 

\newcommand{\tsa}{\underline{t}} 
\newcommand{\tea}{\bar{t}} 
\newcommand{\nva}{\bar{k}} 
\newcommand{\brd}{t^\text{BRD}} 
\newcommand{\trd}{t^\text{RD}} 
\newcommand{\rvs}{\mathcal{N}^\text{RP}} 
\newcommand{\dsen}{d^\text{A}} 
\newcommand{\avoidt}{\mathcal{A}} 

\newtheorem{assumption}{Assumption}

\newtheorem{remark}{Remark}
\newtheorem{observation}{Observation}

\title{\LARGE \bf Safe and Resilient Multi-vehicle Trajectory Planning Under Adversarial Intruder}
\author{Somil Bansal*, Mo Chen*, and Claire J. Tomlin
\thanks{This research is supported by NSF under CPS:ActionWebs (CNS-931843), under the CPS Frontiers VehiCal project (1545126), by the UC-Philippine-California Advanced Research Institute under project IIID-2016-005, and by the ONR MURI Embedded Humans (N00014-16-1-2206).}
\thanks{* Both authors contributed equally to this work. All authors are with the Department of Electrical Engineering and Computer Sciences, University of California, Berkeley. \{somil, mochen72, tomlin\}@eecs.berkeley.edu}
}

\begin{document}
\maketitle
\thispagestyle{empty}
\pagestyle{empty}

\begin{abstract}
Provably safe and scalable multi-vehicle trajectory planning is an important and urgent problem. Hamilton-Jacobi (HJ) reachability is an ideal tool for analyzing such safety-critical systems and has been successfully applied to several small-scale problems. However, a direct application of HJ reachability to multi-vehicle trajectory planning is often intractable due to the ``curse of dimensionality." To overcome this problem, the sequential trajectory planning (STP) method, which assigns strict priorities to vehicles, was proposed; STP allows multi-vehicle trajectory planning to be done with a linearly-scaling computation complexity. However, if a vehicle not in the set of STP vehicles enters the system, or even worse, if this vehicle is an adversarial intruder, the previous formulation requires the entire system to perform replanning, an intractable task for large-scale systems. In this paper, we make STP more practical by providing a new algorithm where replanning is only needed only for a fixed number of vehicles, irrespective of the total number of STP vehicles. Moreover, this number is a design parameter, which can be chosen based on the computational resources available during run time. We demonstrate this algorithm in a representative simulation of an urban airspace environment.    
\end{abstract}

\section{Introduction \label{sec:introduction}}
Recently, there has been an immense surge of interest in the use of unmanned aerial systems (UASs) for civil applications \cite{Tice91, Debusk10, Amazon16, AUVSI16, BBC16}, which will involve unmanned aerial vehicles (UAVs) flying in urban environments, potentially in close proximity to humans, other UAVs, and other important assets. As a result, new scalable ways to organize an airspace are required in which potentially thousands of UAVs can fly together \cite{FAA13, Kopardekar16}.

One essential problem that needs to be addressed for this endeavor to be successful is that of trajectory planning: how a group of vehicles in the same vicinity can reach their destinations while avoiding situations which are considered dangerous, such as collisions. Many previous studies address this problem under different assumptions. In some studies, specific control strategies for the vehicles are assumed, and approaches such as those involving induced velocity obstacles \cite{Fiorini98, Chasparis05, Vandenberg08,Wu2012} and those involving virtual potential fields to maintain collision avoidance \cite{Olfati-Saber2002, Chuang07} have been used. Methods have also been proposed for real-time trajectory generation \cite{Feng-LiLian2002}, for path planning for vehicles with linear dynamics in the presence of obstacles with known motion \cite{Ahmadzadeh2009}, and for cooperative path planning via waypoints which do not account for vehicle dynamics \cite{Bellingham}. Other related work is in the collision avoidance problem without path planning. These results include those that assume the system has a linear model \cite{Beard2003, Schouwenaars2004, Stipanovic2007}, rely on a linearization of the system model \cite{Massink2001, Althoff2011}, assume a simple positional state space \cite{Lin2015}, and many others \cite{Lalish2008, Hoffmann2008, Chen2016}.

However, methods to flexibly plan provably safe and dynamically feasible trajectories without making strong assumptions on the vehicles' dynamics and other vehicles' motion are lacking. Moreover, any trajectory planning scheme that addresses collision avoidance must also guarantee both goal satisfaction and safety of UAVs despite disturbances and communication faults \cite{Kopardekar16}. Furthermore, unexpected scenarios such as UAV malfunctions or even UAVs with malicious intent need to be accounted for. Finally, the proposed scheme should scale well with the number of vehicles.

Hamilton-Jacobi (HJ) reachability-based methods \cite{Barron90, Mitchell05, Bokanowski10, Bokanowski11, Margellos11, Fisac15} are particularly suitable in the context of UAVs because of the formal guarantees provided. In this context, one computes the reach-avoid set, defined as the set of states from which the system can be driven to a target set while satisfying time-varying state constraints at all times. A major practical appeal of this approach stems from the availability of modern numerical tools which can compute various definitions of reachable sets \cite{Sethian96, Osher02, Mitchell02, Mitchell07b}. These numerical tools, for example, have been successfully used to solve a variety of differential games, trajectory planning problems, and optimal control problems \cite{Bayen07, Ding08, Bouffard12, Huang11}. 
However, reachable set computations involve solving a HJ partial differential equation (PDE) or variational inequality (VI) on a grid representing a discretization of the state space, resulting in an \textit{exponential} scaling of computational complexity with respect to the system dimensionality. Therefore, reachability analysis or other dynamic programming-based methods alone are not suitable for managing the next generation airspace, which is a large-scale system with a high-dimensional joint state space because of the possible high density of vehicles that needs to be accommodated \cite{Kopardekar16}.  

To overcome this problem, the priority-based Sequential Trajectory Planning (STP) method has been proposed \cite{Chen15c, Bansal2017}. In this context, higher-priority vehicles plan their trajectories without taking into account the lower-priority vehicles, and lower-priority vehicles treat higher-priority vehicles as moving obstacles. Under this assumption, time-varying formulations of reachability \cite{Bokanowski11, Fisac15} can be used to obtain the optimal and provably safe trajectories for each vehicle, starting from the highest-priority vehicle. Thus, the curse of dimensionality is overcome at the cost of a structural assumption, under which the computation complexity scales just \textit{linearly} with the number of vehicles. In addition, such a structure has the potential to flexibly divide up the airspace for the use of many UAVs and allows tractable multi-vehicle trajectory-planning. Practically, different economic mechanisms can be used to establish a priority order. One example could be first-come-first-serve mechanism, as highlighted in NASA's concept of operations for UAS traffic management \cite{Kopardekar16}.

However, if a vehicle not in the set of STP vehicles enters the system, or even worse, if this vehicle has malicious intent, the original plan can lead to a vehicle colliding with another vehicle, leading to a domino effect, causing the entire STP structure to collapse. Thus, STP vehicles must plan with an additional safety margin that takes a potential intruder into account. The authors in \cite{Chen2016d} propose an STP algorithm that accounts for such a potential intruder. However, a new full-scale trajectory planning problem is required to be solved in real time to ensure safe transit of the vehicles to their respective destinations. Since the replanning must be done in real-time, the proposed algorithm in \cite{Chen2016d} is intractable for large-scale systems even with the STP structure. In this work, we propose a novel algorithm that limits the replanning to a \textit{fixed number of vehicles}, irrespective of the total number of STP vehicles. Moreover, this design parameter can be chosen beforehand based on the computational resources available. 

Intuitively, for every vehicle, we compute a \textit{separation region} such that the vehicle needs to account for the intruder if and only if the intruder is inside this separation region. We then compute a \textit{buffer region} between the separation regions of any two vehicles, and ensure that this buffer is maintained as vehicles are traveling to their destinations. Thus, to intrude every additional vehicle, the intruder must travel through the buffer region. Therefore, we can design the buffer region size such that the intruder can affect at most a specified number of vehicles within some duration. A high-level overview of the proposed algorithm is provided in Algorithm \ref{alg:basic_idea}.    
\begin{algorithm}[tb]
\SetKwInOut{Input}{input}
\SetKwInOut{Output}{output}
	\DontPrintSemicolon
	\caption{Overview of the proposed intruder avoidance algorithm (planning phase)}
	\label{alg:basic_idea}
	\Input{Set of vehicles $\veh_i, i = 1, \ldots, \N$ in the descending priority order;\newline
	Vehicle dynamics and initial states;\newline
	Vehicle destinations and any obstacles to avoid;\newline
	Intruder dynamics;\newline
	$\nva$: Maximum number of vehicles allowed to re-plan their trajectories.}
    \Output{Provably safe vehicle trajectories to respective destinations despite disturbances and intruder;\newline 
    Intruder avoidance and goal-satisfaction controller.}
	\For{\text{$i=1:N$}}{
			compute the separation region of $\veh_i$;\;
			compute the required buffer region based on $\nva$;\;
			use STP algorithm for trajectory planning of $\veh_i$ such that the buffer region is maintained between $\veh_i$ and $\veh_j$ for all $j<i$;\;
			output the trajectory and the optimal controller for $\veh_i$.\;
		}
\end{algorithm}

In Section \ref{sec:formulation}, we formalize the STP problem in the presence of disturbances and adversarial intruders. In Section \ref{sec:background}, we present an overview of time-varying reachability and basic STP algorithms in \cite{Chen15c}, \cite{Bansal2017}. In Section \ref{sec:intruder}, we present our proposed algorithm. Finally, we illustrate this algorithm through a fifty-vehicle simulation in an urban environment in Section \ref{sec:simulations}. Notations are summarized in Table \ref{table:notation}.

\section{Sequential Trajectory Planning Problem \label{sec:formulation}}
Consider $\N$ vehicles $\veh_i, i = 1, \ldots, \N$ (also denoted as \textit{STP vehicles}) which participate in the STP process. We assume their dynamics are given by 

\begin{equation}
\label{eq:dyn}
\begin{aligned}
\dot\state_i &= \fdyn_i(\state_i, \ctrl_i, \dstb_i), t \le \sta_i \\
\ctrl_i &\in \cset_i, \dstb_i \in \dset_i, i = 1 \ldots, \N
\end{aligned}
\end{equation}

\noindent where $\state_i \in \R^{n_i}$, $\ctrl_i \in \cset_i$ and $\dstb_i \in \dset_i$, respectively, represent the state, control and disturbance experienced by vehicle $\veh_i$. We partition the state $\state_i$ into the position component $\pos_i \in \R^{n_\pos}$ and the non-position component $\npos_i \in \R^{n_i - n_\pos}$: $\state_i = (\pos_i, \npos_i)$. 
We will use the sets $\cfset_i, \dfset_i$ to respectively denote the set of functions from which the control and disturbance functions $\ctrl_i(\cdot), \dstb_i(\cdot)$ are drawn.




Each vehicle $\veh_i$ has initial state $\state^0_i$, and aims to reach its target $\targetset_i$ by some scheduled time of arrival $\sta_i$. The target in general represents some set of desirable states, for example the destination of $\veh_i$. 
On its way to $\targetset_i$, $\veh_i$ must avoid a set of static obstacles $\soset_i \subset \R^{n_i}$, which could represent any set of states, such as positions of tall buildings, that are forbidden. Each vehicle $\veh_i$ must also avoid the danger zones with respect to every other vehicle $\veh_j, j\neq i$. For simplicity, we define the danger zone of $\veh_i$ with respect to $\veh_j$ to be
\begin{equation} \label{eqn:danger_zone_defn}
\dz_{ij} = \{(\state_i, \state_j): \|\pos_i - \pos_j\|_2 \le \rc\}
\end{equation}

The danger zones in general can represent any joint configurations between $\veh_i$ and $\veh_j$ that are considered to be unsafe. In particular, $\veh_i$ and $\veh_j$ are said to have collided if $(\state_i, \state_j) \in \dz_{ij}$.

In addition to the obstacles and danger zones, an intruder vehicle $\veh_{\intr}$ may also appear in the system. An intruder vehicle may have malicious intent or simply be a non-participating vehicle that could accidentally collide with other vehicles. This general definition of intruder allows us to develop algorithms that can also account for vehicles who are not communicating with the STP vehicles or do not know about the STP structure. In general, the effect of intruders on vehicles in structured flight can be unpredictable, since the intruders in principle could be adversarial in nature, and the number of intruders could be arbitrary, in which case a collision avoidance problem must be solved for each STP vehicle in the joint state-space of all intruders and the STP vehicle. Therefore, to make our analysis tractable, we make the following two assumptions.

\begin{assumption}
\label{as:avoidOnce}
At most one intruder affects the STP vehicles at any given time. The intruder is removed after a duration of $\iat$. 
\end{assumption}

This assumption can be valid in situations where intruders are rare, and that some fail-safe or enforcement mechanism exists to force the intruder out of the planning space. For example, when STP vehicles are flying at a particular altitude level, the removal of the intruder can be achieved by forcing the intruder to exit the altitude level. Practically, over a large region of the unmanned airspace, this assumption implies that there would be one intruder vehicle per ``planning region''. Each planning region would perform STP independently from the others. One would design planning regions to be an appropriate size such that it is reasonable to assume at most one intruder would appear. The entire large region would be composed of several planning regions.
 
Let the time at which intruder appears in the system be $\tsa$. Assumption \ref{as:avoidOnce} implies that any vehicle $\veh_i$ would need to avoid the intruder $\veh_{\intr}$ for a maximum duration of $\iat$.
Note that we do not pose any restriction on $\tsa$; we only assume that once the intruder appears, it stays for a maximum duration of $\iat$.
\begin{assumption}
\label{as:dynKnown}
The dynamics of the intruder are known and given by $\dot\state_\intr = f_\intr(\state_\intr, \ctrl_\intr, \dstb_\intr)$.
\end{assumption}
Assumption \ref{as:dynKnown} is required for HJ reachability analysis. In situations where the dynamics of the intruder are not known exactly, a conservative model of the intruder may be used instead. We also denote the initial state of the intruder as $\state_{\intr}^0.$ Note that we only assume that the dynamics of the intruder are known, but its initial state $\state_{\intr}^0$, control $\ctrl_\intr$ and disturbance $\dstb_\intr$ it experiences are unknown.

Given the set of STP vehicles, their targets $\targetset_i$, the static obstacles $\soset_i$, the vehicles' danger zones with respect to each other $\dz_{ij}$, and the intruder dynamics $f_\intr(\cdot)$, our goal is as follows. For each vehicle $\veh_i$, synthesize a controller which guarantees that $\veh_i$ reaches its target $\targetset_i$ at or before the scheduled time of arrival $\sta_i$, while avoiding the static obstacles $\soset_i$, the danger zones with respect to all other vehicles $\dz_{ij}, j\neq i$, and the intruder vehicle $\veh_{\intr}$, irrespective of the control strategy of the intruder. In addition, we would like to obtain the latest departure time $\ldt_i$ such that $\veh_i$ can still arrive at $\targetset_i$ on time.

Due to the high dimensionality of the joint state-space, a direct dynamic programming-based solution is intractable. Therefore, the authors in \cite{Chen15c} proposed to assign a priority to each vehicle, and perform STP given the assigned priorities. Without loss of generality, let $\veh_j$ have a higher-priority than $\veh_i$ if $j<i$. Under the STP scheme, higher-priority vehicles can ignore the presence of lower-priority vehicles, and perform trajectory planning without taking into account the lower-priority vehicles' danger zones. A lower-priority vehicle $\veh_i$, on the other hand, must ensure that it does not enter the danger zones of the higher-priority vehicles $\veh_j, j<i$ or the intruder vehicle $\veh_{\intr}$; each higher-priority vehicle $\veh_j$ induces a set of time-varying obstacles $\ioset_i^j(t)$, which represents the possible states of $\veh_i$ such that a collision between $\veh_i$ and $\veh_j$ or $\veh_i$ and $\veh_{\intr}$ could occur.

It is straightforward to see that if each vehicle $\veh_i$ is able to plan a trajectory that takes it to $\targetset_i$ while avoiding the static obstacles $\soset_i$, the danger zones of \textit{higher-priority vehicles} $\veh_j, j<i$, and the danger zone of the \textit{intruder} $\veh_{\intr}$, then the set of STP vehicles $\veh_i, i=1,\ldots,\N$ would all be able to reach their targets safely. Under the STP scheme, trajectory planning can be done sequentially in descending order of vehicle priority in the state space of only a single vehicle. Thus, STP provides a solution whose complexity scales linearly with the number of vehicles. It is important to note that the trajectory planning is always feasible for the lower-priority vehicle under STP because a lower-priority vehicle can always depart early to avoid the higher-priority vehicle on its way to its destination.

When an intruder appears in the system, STP vehicles may need to avoid the intruder to ensure safety. Depending on the initial state of the intruder and its control policy, a vehicle will potentially need to apply different avoidance controls leading to different final states after avoiding the intruder. Therefore, a vehicle's control policy that ensures its successful transit to its destination needs to account for all such possible final states, which is a trajectory planning problem with multiple initial states and a single destination, and is hard to solve in general. Thus, we divide the intruder avoidance problem into two sub-problems: (i) we first design a control policy that ensures a successful transit to the destination if no intruder appears and that successfully avoids the intruder otherwise (Algorithm \ref{alg:basic_idea}). (ii) After the intruder disappears from the system, we replan the trajectories of the affected vehicles. Following the same theme and assumptions, the authors in \cite{Chen2016d} present an algorithm to avoid an intruder in STP formulation; however, in the worst-case, the algorithm might need to replan the trajectories for \textit{all} STP vehicles. Our goal in this work is to present an algorithm that ensures that only a \textit{small and fixed} number of vehicles needs to replan their trajectories, regardless of the total number of vehicles, resulting in a constant replanning time. In particular, we answer the following inter-dependent questions:
\begin{enumerate}
\item How can each vehicle guarantee that it will reach its target set without getting into any danger zones, despite no knowledge of the intruder initial state, the time at which it appears, its control strategy, and disturbances it experiences?
\item How can we ensure that replanning only needs to be done for at most a chosen fixed maximum number of vehicles after the intruder disappears from the system? \label{question2}
\end{enumerate}

\begin{table*}
    \caption{Mathematical notation and their interpretation (in the alphabetical order of symbols).}    
    \resizebox{\hsize}{!}{
    \begin{tabular}{ |>{\centering\arraybackslash}m{1.2cm}| m{5.2cm} | m{2.8cm} | m{0.3cm + \columnwidth} |}
    \hline
    \textbf{Notation} & \textbf{Description} & \textbf{Location} & \textbf{Interpretation} \\ \hline
    
    $\buff_{ij}(t)$ & Buffer region between vehicle $j$ and vehicle $i$ & Beginning of Section \ref{sec:buffRegion_case1} & The set of all possible states for which the separation requirement may be violated between vehicle $j$ and vehicle $i$ for some intruder strategy. If vehicle $i$ is outside this set, then the intruder will need atleast a duration of $\brd$ to go from the avoid region of vehicle $j$ to the avoid region of vehicle $i$.  \\ \hline    
    
    $\dstb_i$ & Disturbance in the dynamics of vehicle $i$ & Beginning of Section \ref{sec:formulation} & -    \\ \hline   
    $\dstb_{\intr}$ & Disturbance in the dynamics of the intruder & Assumption \ref{as:dynKnown} & -    \\ \hline   
    
    $\fdyn_i$ & Dynamics of vehicle $i$ & Beginning of Section \ref{sec:formulation} & -    \\ \hline
    $f_\intr$ & Dynamics of the intruder & Assumption \ref{as:dynKnown} & -    \\ \hline
    $f_r$ & Relative dynamics between two vehicles & Equation \eqref{eq:reldyn} & - \\ \hline
    
    $\obsset_i(t)$ & The overall obstacle for vehicle $i$ & Equation \eqref{eq:obsseti} & The set of states that vehicle $i$ must avoid on its way to the destination. \\ \hline
    
    $\npos_i$ & Non-position state component of vehicle $i$ & Beginning of Section \ref{sec:formulation} & -    \\ \hline
    
    $\nva$ & - & Beginning of Section \ref{sec:intruder} & The maximum number of vehicles that should apply the avoidance maneuver or the maximum number of vehicles that we can replan trajectories for in real-time.    \\ \hline    
    
    $\targetset_i$ & Target set of vehicle $i$ & Beginning of Section \ref{sec:formulation} & The destination of vehicle $i$.    \\ \hline
    
    $\boset_j(t)$ & Base obstacle induced by vehicle $j$ at time $t$ & Equations (25), (31) and (37) in \cite{Chen2016d} & The set of all possible states that vehicle $j$ can be in at time $t$ if the intruder does not appear in the system till time $t$. \\ \hline    
    
    $\N$ & Number of STP vehicles & Beginning of Section \ref{sec:formulation} & -    \\ \hline
    $\rvs$ & - & Equation \eqref{eq:RVS} & The set of vehicles that need to replan their trajectories after the intruder disappears. These are also the set of vehicles that were forced to apply an avoidance maneuver. \\ \hline

	$\ioset_i^j(t)$ & Induced obstacle by vehicle $j$ for vehicle $i$ & After Assumption \ref{as:dynKnown} in Section \ref{sec:formulation} & The possible states of vehicle $i$ such that a collision between vehicle $i$ and vehicle $j$ or vehicle $i$ and the intruder vehicle (if present) could occur.    \\ \hline 
    $\soset_i$ & Static obstacle for vehicle $i$ & Beginning of Section \ref{sec:formulation} & Obstacles that vehicle $i$ needs to avoid on its way to destination, e.g, tall buildings. \\ \hline    
    
    $\pos_i$ & Position of vehicle $i$ & Beginning of Section \ref{sec:formulation} & -    \\ \hline
     
    $\veh_{i}$ & $i$th STP vehicle & Beginning of Section \ref{sec:formulation} & -  \\ \hline
    $\veh_{\intr}$ & The intruder vehicle & Assumption \ref{as:avoidOnce} & -  \\ \hline
    
    $\rc$ & Danger zone radius & Equation \eqref{eqn:danger_zone_defn} & The closest distance between vehicle $i$ and vehicle $j$ that is considered to be safe. \\ \hline  
    
    $\sep_j(t)$ & Separation region of vehicle $j$ at time $t$ & Beginning of Section \ref{sec:sepRegion_case1} & The set of all states of intruder at time $t$ for which vehicle $j$ is forced to apply an avoidance maneuver. \\ \hline        
    
    $\tsa_i$ & Avoid start time of vehicle $i$ & Equation \eqref{eqn:avoidStartTime2} & The first time at which vehicle $i$ is forced to apply an avoidance maneuver by the intruder vehicle. Defined to be $\infty$ if vehicle $i$ never applies an avoidance maneuver.\\ \hline    
    $\brd$ & Buffer region travel duration & Beginning of Section \ref{sec:intruder} & The minimum time required for the intruder to travel through the buffer region between any pair of vehicles. \\ \hline
    $\iat$ & Intruder avoidance time & Assumption \ref{as:avoidOnce} & The maximum duration for which the intruder is present in the system. \\ \hline
    $\tsa$ & Intruder appearance time & After Assumption \ref{as:avoidOnce} & The time at which the intruder appears in the system. \\ \hline
    $\ldt_i$ & Latest departure time of vehicle $i$ & End of Section \ref{sec:formulation} & The latest departure time for vehicle $i$ such that it safely reaches its destination by the scheduled time of arrival. \\ \hline
    $\sta_i$ & Scheduled time of arrival (STA) of vehicle $i$ & Beginning of Section \ref{sec:formulation} & The time by which vehicle $i$ is required to reach its destination.    \\ \hline
    
    $\ctrl_i$ & Control of vehicle $i$ & Beginning of Section \ref{sec:formulation} & -    \\ \hline   
    $\ctrl_{\intr}$ & Control of the intruder & Assumption \ref{as:dynKnown} & -    \\ \hline    
    ${\ctrl^{\text{A}}_{i}}$ & Optimal avoidance control of vehicle $i$ & Equation \eqref{eqn:optAvoidCtrl} & The control that vehicle $i$ need to apply to successfully avoid the intruder once the relative state between vehicle $i$ and the intruder reaches the boundary of the avoid region of vehicle $i$.  \\ \hline 
    ${\ctrl^{\text{PP}}_{i}}$ & Nominal control & Equation \eqref{eqn:PPPolicy} & The nominal control for vehicle $i$ that will ensure its successful transition to its destination if the intruder does not force it to apply an avoidance maneuver. This control law corresponds to the nominal trajectory of vehicle $i$. \\ \hline
	$\ctrl_i^{\text{RP}}$ & The overall controller for vehicle $i$ & Equation \eqref{eqn:full_controller} & The overall controller for vehicle $i$ that will ensure a successful and safe transit to its destination despite the worst-case intruder strategy. \\ \hline

    $\brs^{\text{A}}_{i}(\tau, \iat)$ & Avoid region of vehicle $i$ & Equation \eqref{eqn:avoidBRS} & The set of relative states $\state_{\intr i}$ for which the intruder can force vehicle $i$ to enter in the danger zone $\dz_{i \intr}$ within a duration of $(\iat-\tau)$. \\ \hline

    $\brs^{\text{B}}_{i}(0, \brd)$ & Relative buffer region & Beginning of Section \ref{sec:buffRegion_case1} & The set of all states from which it is possible to reach the boundary of the avoid region of vehicle $i$ within a duration of $\brd$. \\ \hline 
    
   $\brs^{\text{PP}}_{i}$ & - & Equation \eqref{eqn:intrBRS1} & The set of all states that vehicle $i$ needs to avoid in order to avoid a collision with the static obstacles while applying an avoidance maneuver. \\ \hline      
    
   $\brs^{\text{S}}_{i}$ & - & Equation \eqref{eq:ObsBRS_static} & The set of all initial states of vehicle $i$ from which it is guaranteed to safely reach its destination if the intruder does not force it to apply an avoidance maneuver and successfully and safely avoid the intruder in case needs it does. \\ \hline  
   
    $\state_i$ & State of vehicle $i$ & Beginning of Section \ref{sec:formulation} & - \\ \hline
    $\state_{\intr}$ & State of the intruder vehicle & Assumption \ref{as:dynKnown} & - \\ \hline
    $\state_i^0$ & Initial state of vehicle $i$ & Beginning of Section \ref{sec:formulation} & - \\ \hline
    $\state_{\intr}^0$ & Initial state of the intruder vehicle & Assumption \ref{as:dynKnown} & - \\ \hline
    $\state_{\intr i}$ & Relative state between the intruder and vehicle $i$ & Equation \eqref{eq:reldyn} & - \\ \hline
    
    $\dz_{ij}$ & Danger zone between vehicle $i$ and vehicle $j$ & Equation \eqref{eqn:danger_zone_defn} & Set of all states of vehicle $i$ and vehicle $j$ which are within unsafe distance of each other. The vehicles are said to have collided if their states belong to $\dz_{ij}$. \\ \hline
    \end{tabular} }
    \label{table:notation}
\end{table*}

\section{Background \label{sec:background}}
In this section, we first present the basic STP algorithm \cite{Chen15c} in which disturbances and the intruders are ignored and perfect information of vehicles' positions is assumed. We then briefly discuss the different algorithms proposed in \cite{Bansal2017} to account for disturbances in vehicles' dynamics. All of these algorithms use time-varying reachability analysis to provide goal satisfaction and safety guarantees; therefore, we start with an overview of time-varying reachability.

%

\subsection{Time-Varying Reachability Background \label{sec:HJIVI}}
We will be using reachability analysis to compute either a backward reachable set (BRS) $\brs$, a forward reachable set (FRS) $\frs$, or a sequence of BRSs and FRSs, given some target set $\targetset$, time-varying obstacle $\obsset(t)$ which captures trajectories of higher-priority vehicles, and the Hamiltonian function $\ham$ which captures the system dynamics as well as the roles of the control and disturbance. The BRS $\brs$ in a time interval $[t, t_f]$ or FRS $\frs$ in a time interval $[t_0, t]$ will be denoted by $\brs(t, t_f)$ or $\frs(t_0, t)$ respectively. Typically, when computing the BRS, $t_f$ will be some fixed final time, for example the scheduled time of arrival $\sta$. When computing the FRS, $t_0$ will be some fixed initial time, for example the starting time or the present time.

Several formulations of reachability are able to account for time-varying obstacles \cite{Bokanowski11, Fisac15} (or state constraints in general). For our application in STP, we utilize the formulation in \cite{Fisac15}, in which a BRS is computed by solving the following \textit{final value} double-obstacle HJ VI:

\begin{equation}
\label{eq:HJIVI_BRS}
\begin{aligned}
\max \Big\{ \min \{&D_t \valfunc(t, \state) + \ham(t, \state, \nabla \valfunc(t, \state)), \fc(\state) - \valfunc(t, \state) \}, \\
&-\obsfunc(t, \state) - \valfunc(t, \state) \Big\} = 0, \quad t \le t_f \\
&\valfunc(t_f, \state) = \max\{\fc(\state), -\obsfunc(t_f, \state)\}
\end{aligned}
\end{equation}

In a similar fashion, the FRS is computed by solving the following \textit{initial value} HJ PDE:

\begin{equation}
\label{eq:HJIVI_FRS}
\begin{aligned}
D_t \valfuncfwd(t, \state) + &\ham(t, \state, \nabla \valfuncfwd(t, \state)) = 0 , \quad t \ge t_0  \\
&\valfuncfwd(t_0, \state) = \max\{\fc(\state), -\obsfunc(t_0, \state)\}
\end{aligned}
\end{equation}

In both \eqref{eq:HJIVI_BRS} and \eqref{eq:HJIVI_FRS}, the function $\ic(\state)$ is the implicit surface function representing the target set $\targetset = \{\state: \ic(\state) \le 0\}$. Similarly, the function $\obsfunc(t, \state)$ is the implicit surface function representing the time-varying obstacles $\obsset(t) = \{\state: \obsfunc(t,\state)\le 0\}$. The BRS $\brs(t, t_f)$ and FRS $\frs(t_0, t)$ are given by

\begin{equation}
\label{eq:implicitValfuncs}
\begin{aligned}
\brs(t, t_f) &= \{\state: \valfunc(t, \state) \le 0\} \\
\frs(t_0, t) &= \{\state: \valfuncfwd(t, \state) \le 0 \}
\end{aligned}
\end{equation}

Some of the reachability computations will not involve an obstacle set $\obsset(t)$, in which case we can simply set $\obsfunc(t, \state) \equiv \infty$ which effectively means that the outside maximum is ignored in \eqref{eq:HJIVI_BRS}. Also, note that unlike in \eqref{eq:HJIVI_BRS}, there is no inner minimization in \eqref{eq:HJIVI_FRS}. As we will see later, we will be using the BRS to determine all states that can reach some target set \textit{within the time horizon} $[t,t_f]$, whereas we will be using the FRS to determine where a vehicle could be \textit{at some particular time} $t$. 
%
%

The Hamiltonian, $\ham(t, \state, \nabla \valfunc(t,\state))$, depends on the system dynamics, and the role of control and disturbance. Whenever $\ham$ does not depend explicitly on $t$, we will drop $t$ from the argument. In addition, the optimization of Hamiltonian gives the optimal control $\ctrl^*(t, \state)$ and optimal disturbance $\dstb^*(t, \state)$, once $\valfunc$ is determined. For BRSs, whenever the existence of a control (``$\exists \ctrl$'') or disturbance is sought, the optimization is a minimum over the set of controls or disturbance. Whenever a BRS characterizes the behavior of the system for all controls (``$\forall \ctrl$'') or disturbances, the optimization is a maximum. We will introduce precise definitions of reachable sets, expressions for the Hamiltonian, expressions for the optimal controls as needed for the many different reachability calculations we use. 
%
%
%
%
%
%

\subsection{STP Without Disturbances and Intruder\label{sec:basic}}
In this section, we give an overview of the basic STP algorithm assuming that there is no disturbance and no intruder affecting the vehicles. 
The majority of the content in this section is taken from \cite{Chen15c}. 

Recall that among the STP vehicles $\veh_i, i=1,\ldots,N$, $\veh_j$ has a higher priority than $\veh_i$ if $j<i$. In the absence of disturbances, we can write the dynamics of the STP vehicles as
\begin{equation}
\label{eq:dyn_no_dstb}
\begin{aligned}
\dot\state_i = \fdyn_i(\state_i, \ctrl_i), t \le \sta_i,\quad \ctrl_i &\in \cset_i, 
\end{aligned}
\end{equation}

In STP, each vehicle $\veh_i$ plans its trajectory while avoiding static obstacles $\soset_i$ and the obstacles $\ioset_i^j(t)$ induced by higher-priority vehicles $\veh_j, j<i$. Trajectory planning is done sequentially in descending priority, $\veh_1, \veh_2, \ldots, \veh_{\N}$. During its trajectory planning process, $\veh_i$ ignores the presence of lower-priority vehicles $\veh_k, k>i$.
From the perspective of $\veh_i$, each of the higher-priority vehicles $\veh_j, j<i$ induces a time-varying obstacle denoted $\ioset_i^j(t)$ that $\veh_i$ needs to avoid. Therefore, each vehicle $\veh_i$ must plan its trajectory while avoiding the union of all the induced obstacles as well as the static obstacles. Let $\obsset_i(t)$ be the union of all the obstacles that $\veh_i$ must avoid on its way to $\targetset_i$:
\begin{equation}
\label{eq:obsseti}
\obsset_i(t)  = \soset_i \cup \bigcup_{j=1}^{i-1} \ioset_i^j(t)
\end{equation}

With full position information of higher-priority vehicles, the obstacle induced for $\veh_i$ by $\veh_j$ is simply
\begin{equation}
\label{eq:ioset_basic}
\ioset_i^j(t) = \{\state_i: \|\pos_i - \pos_j(t)\|_2 \le \rc \}
\end{equation}

Each higher-priority vehicle $\veh_j$ ignores $\veh_i$. Since trajectory planning is done sequentially in descending order of priority, the vehicles $\veh_j, j<i$ would have planned their trajectories before $\veh_i$ does. Thus, in the absence of disturbances, $\pos_j(t)$ is \textit{a priori} known, and therefore $\ioset_i^j(t), j<i$ are known, deterministic moving obstacles, which means that $\obsset_i(t)$ is also known and deterministic. Therefore, the trajectory planning problem for $\veh_i$ can be solved by first computing the BRS $\brs_i^\text{basic}(t, \sta_i)$, defined as follows:
\begin{equation}
\label{eq:BRS_basic}
\begin{aligned}
\brs_i^\text{basic}(t, \sta_i) = & \{y: \exists \ctrl_i(\cdot) \in \cfset_i, \state_i(\cdot) \text{ satisfies \eqref{eq:dyn_no_dstb}}, \\
& \forall s \in [t, \sta_i],\state_i(s) \notin \obsset_i(s), \\
& \exists s \in [t, \sta_i], \state_i(s) \in \targetset_i, \state_i(t) = y\}
\end{aligned}
\end{equation}

The BRS $\brs(t, \sta_i)$ can be obtained by solving \eqref{eq:HJIVI_BRS} with $\targetset = \targetset_i$, $\obsset(t) = \obsset_i(t)$, and the Hamiltonian 
\begin{equation}
\label{eq:basicham}
\ham_i^\text{basic}(\state_i, \costate) = \min_{\ctrl_i\in\cset_i} \costate \cdot \fdyn_i(\state_i, \ctrl_i)
\end{equation}

The optimal control for reaching $\targetset_i$ while avoiding $\obsset_i(t)$ is then given by
\begin{equation}
\label{eq:basicOptCtrl}
\ctrl_i^\text{basic}(t, \state_i) = \arg \min_{\ctrl_i\in\cset_i} \costate \cdot \fdyn_i(\state_i, \ctrl_i)
\end{equation}
\noindent from which the trajectory $\state_i(\cdot)$ can be computed by integrating the system dynamics, which in this case are given by \eqref{eq:dyn_no_dstb}. In addition, the latest departure time $\ldt_i$ can be obtained from the BRS $\brs(t, \sta_i)$ as $\ldt_i = \arg \sup_t \{\state_i^0 \in \brs(t, \sta_i)\}$. The basic STP algorithm is summarized in Algorithm \ref{alg:basic}.
\begin{algorithm}[tb]
\SetKwInOut{Input}{input}
\SetKwInOut{Output}{output}
	\DontPrintSemicolon
	\caption{STP algorithm in the absence of disturbances and intruders}
	\label{alg:basic}
	\Input{STP vehicles $\veh_i$, their dynamics \eqref{eq:dyn_no_dstb}, initial states $\state_i^0$,	destinations $\targetset_i$, static obstacles $\soset_i$}
    \Output{Provably safe trajectories to destinations and goal-satisfaction controllers $\ctrl^\text{basic}(\cdot)$}
	\For{\text{$i=1:N$}}{
			\textbf{Trajectory planning for $\veh_{i}$} \;
			compute the total obstacle set $\obsset_i(t)$ given by \eqref{eq:obsseti}. If $i=1$, $\obsset_i(t) = \soset_i ~ \forall t$;\;
			compute the BRS $\brs_i^\text{basic}(t, \sta_i)$ defined in \eqref{eq:BRS_basic};\;
			\textbf{Trajectory and controller of $\veh_{i}$} \;
			compute the optimal controller $\ctrl_i^\text{basic}(\cdot)$ given by \eqref{eq:basicOptCtrl};\;
			determine the trajectory $\state_i(\cdot)$ using vehicle dynamics \eqref{eq:dyn_no_dstb} and the control $\ctrl_i^\text{basic}(\cdot)$; \;
			output the trajectory and optimal controller for $\veh_i$.\;
			\textbf{Obstacles induced by $\veh_{i}$} \;
			given the trajectory $\state_i(\cdot)$, compute the induced obstacles $\ioset_k^i(t)$ given by \eqref{eq:ioset_basic} for all $k>i$.
		}
\end{algorithm}
%
\subsection{STP With Disturbances and Without Intruder\label{sec:distb}}
Disturbances and incomplete information significantly complicate the STP scheme. The main difference is that the vehicle dynamics satisfy \eqref{eq:dyn} as opposed to \eqref{eq:dyn_no_dstb}. Committing to exact trajectories is therefore no longer possible, since the disturbance $d_i(\cdot)$ is \textit{a priori} unknown. Thus, the induced obstacles $\ioset_i^j(t)$ are no longer just the danger zones centered around positions, unlike in \eqref{eq:ioset_basic}. In particular, a lower-priority vehicle needs to account for all possible states that the higher-priority vehicles could be in. To do this, the lower-priority vehicle needs to have some knowledge about the control policy used by each higher-priority vehicle. Three different methods are presented in \cite{Bansal2017} to address the above issues. The methods differ in terms of control policy information that is known to a lower-priority vehicle.
\begin{itemize}
\item \textbf{Centralized control}: A specific control strategy is enforced upon a vehicle; this can be achieved, for example, by some central agent such as an air traffic controller. 
\item \textbf{Least restrictive control}: A vehicle is required to arrive at its targets on time, but has no other restrictions on its control policy. When the control policy of a vehicle is unknown, the least restrictive control can be safely assumed by lower-priority vehicles.
\item \textbf{Robust trajectory tracking}: A vehicle declares a nominal trajectory which can be robustly tracked under disturbances.
\end{itemize}
In each case, a vehicle $\veh_i$ can compute all possible states $\ioset_i^j(t)$ that a higher-priority vehicle $\veh_j$ can be in based on the control strategy information known to the lower priority vehicle. A collision avoidance between $\veh_i$ and $\veh_j$ is thus ensured. We refer to the obstacle $\ioset_i^j(t)$, induced in the presence of disturbances but in the absence of intruders, as \textit{base obstacle} and denote it as $\boset_j(t)$ from here on. Further details of each algorithm are presented in \cite{Bansal2017}.


\section{Response to Intruders \label{sec:intruder}}
In this section, we propose a method to allow vehicles to avoid an intruder while maintaining the STP structure.
Our goal is to design a control policy for each vehicle that ensures separation with the intruder and other STP vehicles, and ensures a successful transit to the destination. 

As discussed in Section \ref{sec:formulation}, depending on the initial state of the intruder and its control policy, a vehicle may arrive at different states after avoiding the intruder. To make sure that the vehicle still reaches its destination, a replanning of vehicle's trajectory is required. Since the replanning must be done in real-time, we also need to ensure that only a small number of vehicles require replanning. In this work, a novel intruder avoidance algorithm is proposed, which will need to replan trajectories only for a \textit{small fixed} number of vehicles, irrespective of the total number of STP vehicles. Moreover, this number is a design parameter, which can be chosen based on the resources available during run time. 

Let $\nva$ denote the maximum number of vehicles that we can replan the trajectories for in real-time. Also, let $\brd = \frac{\iat}{\nva}$. We divide our algorithm in two parts: the planning phase and the replanning phase. In the planning phase, our goal is to divide the flight space of vehicles such that at any given time, any two vehicles are far enough from each other so that an intruder needs at least a duration of $\brd$ to travel from the vicinity of one vehicle to that of another. Since the intruder is present for a total duration of $\iat$, this division ensures that it can only affect at most $\nva$ vehicles despite its best efforts. In particular, we compute a separation region for each vehicle such that the vehicle needs to account for the intruder if and only if the intruder is inside this separation region. We then compute a buffer region between the separation regions of any two vehicles such that the intruder requires atleast a duration of $\brd$ to travel through this region. A high-level overview of the planning phase is presented in Algorithm \ref{alg:basic_idea}. The planning phase ensures that after the intruder disappears, \textit{at most $\nva$} vehicles have to replan their trajectories. In the replanning phase, we re-plan the trajectories of affected vehicles so that they reach their destinations safely. 

Note that our theory assumes worst-case scenarios in terms of the behavior of the intruder, the effect of disturbances, and the planned trajectories of each STP vehicle. This way, we are able to guarantee safety and goal satisfaction of all vehicles in all possible scenarios given the bounds on intruder dynamics and disturbances. To achieve denser operation of STP vehicles, known information about the intruder, disturbances, and specifies of STP vehicle trajectories may be incorporated; however, these considerations are out of the scope of this paper.

The rest of the section is organized as follows. In Sections \ref{sec:intruder_avoid}, we discuss the intruder avoidance control that a vehicle needs to apply within the separation region. In Sections \ref{sec:case1} and \ref{sec:case2_maintext}, we compute the separation and buffer regions for vehicles. 
Trajectory planning that maintains the buffer region between every pair of vehicles is discussed in Section \ref{sec:path_planning}. Finally, the replanning of the trajectories of the affected vehicles is discussed in Section \ref{sec:replan}. 
\subsection{Optimal Avoidance Controller} \label{sec:intruder_avoid}
In this section, our goal is to compute the control policy that a vehicle $\veh_i$ can use to avoid entering in the danger region $\dz_{i\intr}$. We also compute the set of states from which the joint states of $\veh_{\intr}$ and $\veh_i$ can enter the danger zone $\dz_{i\intr}$ despite the best efforts of $\veh_i$ to avoid $\veh_{\intr}$, which is then used to compute the separation region of $\veh_i$ in Section \ref{sec:sepRegion_case1}. 

We define relative dynamics of the intruder $\veh_{\intr}$ with state $\state_\intr$ with respect to $\veh_i$ with state $\state_i$:
\begin{equation}
\label{eq:reldyn}
\begin{aligned}
\state_{\intr i} = \state_\intr - \state_i, \qquad \dot \state_{\intr i} = f_r(\state_{\intr i}, \ctrl_i, \ctrl_\intr, \dstb_i, \dstb_\intr)
\end{aligned}
\end{equation}
Given the relative dynamics, the set of states from which the joint states of $\veh_{\intr}$ and $\veh_{i}$ can enter danger zone $\dz_{i\intr}$ in a duration of $\iat$ despite the best efforts of $\veh_i$ to avoid $\veh_{\intr}$ is given by the backward reachable set $\brs^{\text{A}}_i(\tau, \iat),~ \tau \in [0, \iat]$:
\begin{equation} \label{eqn:avoidBRS}
\begin{aligned}
\brs^{\text{A}}_{i}(\tau, \iat) = & \{y: \forall \ctrl_i(\cdot) \in \cfset_i, \exists \ctrl_\intr(\cdot) \in \cfset_\intr, \exists \dstb_i(\cdot) \in \dfset_i, \\
& \exists \dstb_\intr(\cdot) \in \dfset_\intr, \state_{\intr i}(\cdot) \text{ satisfies \eqref{eq:reldyn}},\\
& \exists s \in [\tau, \iat], \state_{\intr i}(s) \in \targetset^{\text{A}}_{i}, \state_{\intr i}(\tau) = y\},\\
\targetset^{\text{A}}_{i} = & \{\state_{\intr i}: \|\pos_{\intr i}\|_2 \le \rc\}.
\end{aligned}
\end{equation}
The Hamiltonian to compute $\brs^{\text{A}}_{i}(\tau, \iat)$ is given as:
\begin{equation}
H^{\text{A}}_{i}(\state_{\intr i}, \costate) = \max_{\ctrl_i \in \cset_i} \min_{\substack{\ctrl_\intr \in \cset_\intr, \\ \dstb_\intr \in \dset_\intr, \\ \dstb_i \in \dset_i}} \costate \cdot f_r(\state_{\intr i}, \ctrl_i, \ctrl_\intr, \dstb_i, \dstb_\intr).
\end{equation}
We refer to $\brs^{\text{A}}_i(\tau, \iat)$ as \textit{avoid region} from here on. The interpretation of $\brs^{\text{A}}_{i}(\tau, \iat)$ is that if $\veh_i$ starts inside this set, i.e., $\state_{\intr i}(t) \in  \brs^{\text{A}}_{i}(\tau, \iat)$, then the intruder can force $\veh_i$ to enter the danger zone $\dz_{i\intr}$ within a duration of $(\iat-\tau)$, regardless of the control applied by the vehicle. If $\veh_i$ starts at the boundary of this set (denoted as $\partial \brs^{\text{A}}_{i}(\tau, \iat)$), i.e., $\state_{\intr i}(t) \in  \partial \brs^{\text{A}}_{i}(\tau, \iat)$, it can \textit{barely} successfully avoid the intruder for a duration of $(\iat-\tau)$ using the optimal avoidance control ${\ctrl^{\text{A}}_{i}}$ (referred to as \textit{avoidance maneuver} from here on)
\begin{equation} \label{eqn:optAvoidCtrl}
{\ctrl^{\text{A}}_{i}} = \arg \max_{\ctrl_i \in \cset_i} \min_{\substack{\ctrl_\intr \in \cset_\intr, \\ \dstb_\intr \in \dset_\intr, \\ \dstb_i \in \dset_i}} \costate \cdot f_r(\state_{\intr i}, \ctrl_i, \ctrl_\intr, \dstb_i, \dstb_\intr).
\end{equation}
\noindent Finally, if $\veh_i$ starts outside this set, i.e., $\state_{\intr i}(t) \in \left( \brs^{\text{A}}_{i}(\tau, \iat)\right)^C$, then  $\veh_i$ and $\veh_{\intr}$ cannot instantaneously enter the danger zone $\dz_{i\intr}$, irrespective of the control applied by them at time $t$. In fact, $\veh_i$ can safely apply \textit{any} control as long as it is outside the boundary of this set, but will have to apply the avoidance maneuver to avoid the intruder once it reaches the boundary.

In the worst case, $\veh_i$ might need to avoid the intruder for a duration of $\iat$ starting at $t = \tsa$; thus, the least we must have is that $\state_{\intr i}(\tsa) \in \left(\brs^{\text{A}}_{i}(0, \iat)\right)^C$ to ensure successful avoidance. Otherwise, regardless of what control a vehicle applies, the intruder can force it to enter the danger zone $\dz_{i\intr}$.
\begin{assumption}
\label{as:detection_range}
$\state_{\intr i}(\tsa) \in \left(\brs^{\text{A}}_{i}(0, \iat)\right)^C \forall i \in \{1, \ldots, \N\}$.
\end{assumption}
Intuitively, assumption \ref{as:detection_range} enforces a condition on the detection of the intruder by STP vehicles. For example, if STP vehicles are equipped with circular sensors, then assumption \ref{as:detection_range} implies that STP vehicle must be able to detect a intruder that is within a distance of $\dsen$, where
\begin{equation} \label{eqn:sen_distance}
\dsen = \max\{ \|p_i\|_2: \exists \npos_i, (p_i, \npos_i) \in \brs^{\text{A}}_{i}(0, \iat) \};
\end{equation} 
otherwise, there exists an intruder control strategy such that $\veh_i$ and $\veh_{\intr}$ will collide irrespective of the control used by $\veh_i$. Thus, $\dsen$ is the \textit{minimum} detection range required by any trajectory-planning algorithm to ensure a successful intruder avoidance for all intruder strategies. In general, assumption \ref{as:detection_range} is required to ensure that the intruder gives the STP vehicles ``a chance" to react and avoid it. Hence, for analysis to follow, we assume that assumption \ref{as:detection_range} holds. 

Note that although \eqref{eqn:optAvoidCtrl} gives us a provably successful avoidance control for avoiding the intruder if $\state_{\intr i}(\tsa) \in \left(\brs^{\text{A}}_{i}(0, \iat)\right)^C$, the vehicle may not be able to apply this control because it may lead to a collision with other STP vehicles. Thus, in general, assumption \ref{as:detection_range} is \textit{only necessary not sufficient} to guarantee intruder avoidance. However, we ensure that the STP vehicles are always separated enough from each other so that any vehicle can apply avoidance maneuver if need be. Thus, for the proposed algorithm, assumption \ref{as:detection_range} will also be sufficient for a successful intruder avoidance.
\subsection{Separation and Buffer Regions - Case 1} \label{sec:case1}
In the next two sections, our goal is to compute separation and buffer regions for STP vehicles so that 
at most $\nva$ vehicles are \textit{forced} to apply an avoidance maneuver during the duration $[\tsa, \tsa+\iat]$. 

Intuitively, we divide the duration of $\iat$ into $\nva$ intervals and ensure that atmost one vehicle can be forced to apply the avoidance maneuver during each interval. Thus, by construction, it is guaranteed that at most $\nva$ vehicles apply the avoidance maneuver during the time interval $[\tsa, \tsa+\iat]$. We refer to this structure as the \textit{separation requirement} from here on. To ensure this requirement, we use the reachability theory to find the set of all states of $\veh_i$ such that the separation requirement can be violated between the vehicle pair $(\veh_i, \veh_j)$, $j<i$, at time $t$. During the trajectory planning of $\veh_i$, we then ensure that the vehicle is not in one of these states at time $t$ by using this set of states as ``obstacle". The sequential trajectory planning will therefore guarantee that the separation requirement holds for every STP vehicle pair.

Mathematically, we define
\begin{equation} \label{eqn:avoidStartTime}
\avoidt_{i} := \{t: \state_{\intr i}(t) \in \partial\brs^{\text{A}}_{i}(t-\tsa, \iat), t \in [\tsa, \tsa+\iat]\},
\end{equation}
and the \textit{avoid start time} $\tsa_i$
\begin{equation} \label{eqn:avoidStartTime2}
\tsa_i  = 
\left \{ 
\begin{array}{ll}
\min_{t \in  \avoidt_{i}} t & \mbox{if~} \avoidt_{i} \neq \emptyset\\
\infty & \mbox{otherwise}
\end{array}
\right.
\end{equation}  
By definition of $\brs^{\text{A}}_{i}(\cdot, \iat)$, $\avoidt_{i}$ is the set of all times at which $\veh_i$ must apply an avoidance maneuver and $\tsa_i$ denotes the \textit{first} such time. The separation requirement can thus be written as
\begin{equation} \label{eqn:sep_cond}
\forall i \neq j, \min(\tsa_i, \tsa_j)< \infty \implies |\tsa_i - \tsa_j| \geq \brd := \frac{\iat}{\nva}.
\end{equation}
The separation requirement essentially implies that the intruder requires a time duration of atleast $\brd$ before it can force any additional vehicle to apply an avoidance maneuver. Thus, any two STP vehicles should be ``separated" enough from each other at any given time for such a requirement to hold. 

We now focus on finding what that ``separation set" (also referred to as the buffer region) between $(\veh_i, \veh_j)$. Since the path planning is done in a sequential order, we assume that we have already planned a path for $\veh_j$ and compute the buffer region that $\veh_i$ needs to maintain to ensure that the separation requirement is satisfied.
For this computation, it is sufficient to consider the following two mutually exclusive and exhaustive cases: 
\begin{enumerate}
\item Case 1: $\tsa_j \leq \tsa_i, \tsa_j < \infty$
\item Case 2: $\tsa_i < \tsa_j, \tsa_i <\infty$
\end{enumerate}
In this section, we consider Case 1. Case 2 is discussed in the next section.  

In Case 1, the intruder forces $\veh_j$, the higher-priority vehicle, to apply avoidance control before or at the same time as $\veh_i$, the lower-priority vehicle. 
To ensure the separation requirement in this case, we begin with the following observation which narrows down the intruder scenarios that we need to consider:
\begin{observation} \label{obs1_case1}
Without loss of generality, we can assume that the intruder \textit{appears} in the system at the boundary of the avoid region of $\veh_j$, i.e., $\state_{\intr j}(\tsa) \in \partial \brs^{\text{A}}_{j}(0, \iat)$. Equivalently, we can assume that $\tsa_j = \tsa$.
\end{observation}
Since $\tsa_j \leq \tsa_i$, $\veh_{\intr}$ reaches the boundary of the avoid region of $\veh_j$ before it reaches the boundary of the avoid region of $\veh_i$. Furthermore, by the definition of the avoid region, vehicles $\veh_j$ and $\veh_i$ need not account for the intruder until it reaches the boundary of the avoid region of $\veh_j$. Thus, it is sufficient to consider the worst case $\tsa_j = \tsa$. 
\subsubsection{Separation region} \label{sec:sepRegion_case1}
Recall that the separation region denotes the set of states of the intruder for which a vehicle is forced to apply an avoidance maneuver. In this section our goal is to find $\sep_j(\tsa_j)$, the separation region of $\veh_j$ at the avoid start time. By virtue of Observation \ref{obs1_case1}, we will use $\tsa_j$ and $\tsa$ interchangeably here on.
%

As discussed in Section \ref{sec:intruder_avoid}, $\veh_j$ needs to apply avoidance maneuver at time $\tsa$ only if $\state_{\intr j}(\tsa) \in \partial \brs^{\text{A}}_{j}(0, \iat)$. To compute set $\sep_j(\tsa)$, we thus need to translate these relative states to a set in the state space of the intruder. Therefore, if all possible states of $\veh_j$ at time $\tsa$ are known, then $\sep_j(\tsa)$ can be trivially computed.

Recall from Section \ref{sec:distb} that the base obstacle $\boset_j(t)$ at time $t$ represents all possible states of $\veh_j$ at time $t$, if the intruder doesn't appear in the system until that time. This is precisely the set that we are interested in to compute the separation region.
%
Depending on the information known to a lower-priority vehicle $\veh_i$ about $\veh_j$'s control strategy, we can use one of the three methods described in Section 5 in \cite{Chen2016d} (and Section \ref{sec:distb} of this paper) to compute the base obstacles $\boset_j(\tsa)$. In particular, the base obstacles are respectively given by equations (25), (31) and (37) in \cite{Chen2016d} for the centralized control, the least restrictive control and the robust trajectory tracking algorithms (the three proposed algorithms to account for disturbances in STP). We will explain the computation of the base obstacles further in Section \ref{sec:path_planning}.

Given $\boset_j(\tsa)$, $\sep_j(\tsa)$ can be obtained as:
\begin{equation} \label{eqn:sepRegion_case1}
\sep_j(\tsa) = \boset_j(\tsa) + \partial \brs^{\text{A}}_{j}(0, \iat), ~\tsa \in \R,
\end{equation}
where the ``$+$'' in \eqref{eqn:sepRegion_case1} denotes the Minkowski sum. Since $\sep_j(\tsa)$ represents the set of all states of $\veh_\intr$ for which $\veh_j$ must apply an avoidance maneuver, Observation \ref{obs1_case1} implies that it is sufficient to consider the scenarios where $\state_{\intr}^0 := \state_{\intr}(\tsa) \in \sep_j(\tsa)$.
\subsubsection{Buffer Region} \label{sec:buffRegion_case1}
We are now ready to compute the buffer region $\buff_{ij}$, the set of states for which the separation requirement can be violated between the vehicle pair $(\veh_i, \veh_j)$. Conversely, if $\veh_i$ is outside the buffer region, the separation requirement is satisfied between $(\veh_i, \veh_j)$. We start with recalling the following three results/facts
\begin{enumerate}
\item the separation requirement is equivalent to $|\tsa_i - \tsa| \geq \brd$ (see \ref{eqn:sep_cond})
\item $\state_{\intr}(\tsa) \in \sep_j(\tsa)$ (Section \ref{sec:sepRegion_case1})
\item $\state_{\intr i}(\tsa_i) \in \partial \brs^{\text{A}}_{i}(\tsa_i-\tsa, \iat)$ (Definition of $\tsa_i$).
\end{enumerate}

To compute $\buff_{ij}$, we first compute $\brs^{\text{B}}_{i}(0, \brd)$, the set of all states $\state_{\intr i}$ that can reach the set $\brs^{\text{A}}_{i}(\brd, \iat)$ within a duration of $\brd$. Ensuring that the intruder is outside this set at time $\tsa$ guarantees that it will need a duration of atleast $\brd$ before it can force $\veh_i$ to apply an avoidance maneuver (equivalently, reach the avoid region of $\veh_i$). Since the possible states of the intruder at $\tsa$ is given by $\sep_j(\tsa)$, we thus simply need to ensure that $\sep_j(\tsa)$ is outside $\brs^{\text{B}}_{i}$. Thus, the minimum buffer region is given by:
\begin{equation} \label{eqn:buffRegion_case1}
\buff_{ij}(\tsa) = \sep_j(\tsa) + \brs^{\text{B}}_{i}(0, \brd).
\end{equation} 

We refer to $\brs^{\text{B}}_{i}$ as the \textit{relative buffer region} here on, which is given by the following BRS:
\begin{equation} \label{eqn:buffBRS_case1}
\begin{aligned}
\brs^{\text{B}}_{i}(0, \brd) = & \{y: \exists \ctrl_i(\cdot) \in \cfset_i, \exists \ctrl_\intr(\cdot) \in \cfset_\intr, \exists \dstb_i(\cdot) \in \dfset_i, \\
& \exists \dstb_\intr(\cdot) \in \dfset_\intr, \state_{i \intr}(\cdot) \text{ satisfies \eqref{eq:reldyn}},\\
& \exists s \in [0, \brd], \state_{i \intr}(s) \in -\brs^{\text{A}}_{i}(\brd, \iat),\\
& \state_{i \intr}(t) = y\},\\
-\brs^{\text{A}}_{i}(\brd, \iat) = & \{y: -y \in \brs^{\text{A}}_{i}(\brd, \iat) \}.
\end{aligned}
\end{equation}
The Hamiltonian to compute $\brs^{\text{B}}_{i}(0, \brd)$ is given by:
\begin{equation} 
H^{\text{B}}_{i}(\state_{i \intr}, \costate) = \min_{\substack{\ctrl_i \in \cset_i, \ctrl_\intr \in \cset_\intr, \\ \dstb_i \in \dset_i, \dstb_\intr \in \dset_\intr}} \costate \cdot f_r(\state_{i \intr}, \ctrl_i, \ctrl_\intr, \dstb_i, \dstb_\intr).
\end{equation}
Intuitively, $\brs^{\text{B}}_{i}(0, \brd)$ represents the set of all relative states $\state_{i \intr}$ from which it is possible to reach the boundary of $\brs^{\text{A}}_{i}(\brd, \iat)$ within a duration of $\brd$. Note that we use $-\brs^{\text{A}}_{i}(0, \iat)$ instead of $\brs^{\text{A}}_{i}(0, \iat)$ as the target set for our computation above because the BRS $\brs^{\text{B}}_{i}(0, \brd)$ is computed using the relative state $\state_{i \intr}$ (and not $\state_{\intr i}$.)  
%

To summarize, we can ensure that $(\tsa_i - \tsa_j) \geq \brd$ as long as $\state_i(\tsa) \in \left(\buff_{ij}(\tsa)\right)^C$. This will be ensured by using $\buff_{ij}$ as an obstacle during the path planning of $\veh_i$ (see Section \ref{sec:path_planning}). Consequently, $\veh_{\intr}$ can force at most $\nva$ vehicles to apply an avoidance maneuver during a duration of $\iat$.
\subsubsection{Obstacle Computation} \label{sec:intruderObs_case1}
In Sections \ref{sec:sepRegion_case1} and \ref{sec:buffRegion_case1}, we computed a buffer region between $\veh_i$ and $\veh_j$ such that the separation requirement is satisfied.
However, it still needs to be ensured that a vehicle does not collide with other vehicles while applying an avoidance maneuver. 
In this section, we find the set of states that $\veh_i$ needs to avoid to avoid accidentally entering in $\dz_{ij}$ during an avoidance maneuver. Since the trajectory planning is done in a sequential fashion, being a lower priority vehicle, $\veh_i$ also needs to avoid the states that can lead it to $\dz_{ij}$ while $\veh_j$ is avoiding the intruder.
These sets of states are then used as obstacles during the path planning of $\veh_i$, which ensures that it never enters these ``potentially unsafe" states.

To find this obstacle set, we consider the following two exhaustive cases:
\begin{enumerate}
\item Case A: The intruder affects $\veh_j$, but not $\veh_i$, i.e., $\tsa_j < \infty$ and $\tsa_i = \infty$.
\item Case B: The intruder first affects $\veh_j$ and then $\veh_i$, i.e., $\tsa_j, \tsa_i < \infty$.
\end{enumerate}
For each case, we compute the set of states that $\veh_i$ needs to avoid at time $t$ to avoid entering in $\dz_{ij}$ eventually. Let $_1^A\ioset_i^j(\cdot)$ and $_1^B\ioset_i^j(\cdot)$ denote the corresponding sets of ``obstacles" for the two cases. We begin with the following observation: 
\begin{observation} \label{obs1_case1Obs}
To compute obstacles at time $t$, it is sufficient to consider the scenarios where $\tsa \in [t-\iat, t]$. This is because if $\tsa < t - \iat$, then $\veh_j$ and/or $\veh_i$ will already be in the replanning phase at time $t$ (see assumption \ref{as:avoidOnce}) and hence the two vehicles cannot be in conflict at time $t$. On the other hand, if $\tsa > t$, then $\veh_j$ wouldn't apply any avoidance maneuver at time $t$. 
\end{observation}

\begin{itemize}[leftmargin=*] 
\item \label{sec:intruderObs_case1_caseA} Case A: In this case, only $\veh_j$ applies an avoidance maneuver; therefore, $\veh_i$ should avoid the set of states that can lead to a collision with $\veh_j$ at time $t$ while $\veh_j$ is applying an avoidance maneuver. Note that since $\tsa_j = \tsa$ (by Observation \ref{obs1_case1}), $_1^A\ioset_i^j(t)$ is given by the states that $\veh_j$ can reach while avoiding the intruder, starting from some state in the base obstacle, $\boset_j(\tsa), \tsa \in [t-\iat, t]$. These states can be obtained by computing a FRS from the base obstacles.
\begin{equation} \label{eq:ObsFRS_case1_caseA}
\begin{aligned}
& \frs_{j}^{\mathcal{O}}(\tsa, t) = \{y: \exists \ctrl_j(\cdot) \in \cfset_j, \exists \dstb_j(\cdot) \in \dfset_j, \\
& \quad \state_j(\cdot) \text{ satisfies \eqref{eq:dyn}}, \state_j(\tsa) \in \boset_j(\tsa), \state_j(t) = y\}.
\end{aligned}
\end{equation}
$\frs_{j}^{\mathcal{O}}(\tsa, t)$ represents the set of all possible states that $\veh_j$ can reach after a duration of $(t-\tsa)$ starting from inside $\boset_j(\tsa)$. This FRS can be obtained by solving the HJ VI in \eqref{eq:HJIVI_FRS} with the following Hamiltonian:
\begin{equation}
\ham_{j}^{\mathcal{O}}(\state_j, \costate) = \max_{\ctrl_j \in \cset_j} \max_{\dstb_j \in \dset_j} \costate \cdot f_j (\state_j, \ctrl_j, \dstb_j).
\end{equation} 
Since $\tsa \in [t-\iat, t]$, the induced obstacles in this case can be obtained as:
\begin{equation} \label{eq:intruderObs_case1_caseA} 
\begin{aligned}
_1^A\ioset_i^j(t) & = \{\state_i: \exists y \in \pfrs_j(t), \|\pos_i - y\|_2 \le \rc \}\\
\pfrs_j(t) & = \{p_j: \exists \npos_j, (p_j, \npos_j) \in \bigcup_{\tsa \in [t-\iat, t]} \frs_{j}^{\mathcal{O}}(\tsa, t) \}
\end{aligned}
\end{equation}

\begin{observation} \label{obs1_case1_caseA}
Since the base obstacles represent all possible states of a vehicle in the absence of an intruder, the base obstacle at any time $\tau_2$ is contained within the FRS of the base obstacle at any earlier time $\tau_1 < \tau_2$, computed forward for a duration of $(\tau_2-\tau_1).$ That is, $\boset_j(\tau_2) \subseteq \frs_{j}^{\mathcal{O}}(\tau_1, \tau_2)$, where $\frs_{j}^{\mathcal{O}}(\tau_1, \tau_2)$, as before, denotes the FRS of $\boset_j(\tau_1)$ computed forward for a duration of $(\tau_2-\tau_1)$. The same argument can be applied to the FRSs computed from two different base obstacles $\boset_j(\tau_2)$ and $\boset_j(\tau_1)$, i.e., $\frs_{j}^{\mathcal{O}}(\tau_2, \tau_3) \subseteq \frs_{j}^{\mathcal{O}}(\tau_1, \tau_3)$ if $\tau_1 < \tau_2 < \tau_3$.
\end{observation}

Using observation \ref{obs1_case1_caseA}, $\pfrs_j(t)$ in \eqref{eq:intruderObs_case1_caseA} can be equivalently written as
\begin{equation} \label{eq:intruderObs_case1_caseA_inter}
\pfrs_j(t) = \{p_j: \exists \npos_j, (p_j, \npos_j) \in \frs_{j}^{\mathcal{O}}(t-\iat, t) \}.
\end{equation}

\item \label{sec:intruderObs_case1_caseB} Case B: In this case, first $\veh_j$ applies an avoidance maneuver followed by $\veh_i$. Once $\veh_j$ starts applying avoidance control at time $\tsa = \tsa_j$, it might deviate from its pre-planned control strategy. From the perspective of $\veh_i$, $\veh_j$ can apply any control during $[\tsa, \tsa+\iat]$. Furthermore, $\veh_i$ itself must apply avoidance maneuver during $[\tsa_i, \tsa+\iat]$. Thus, the main challenge in this case is to ensure that $\veh_i$ and $\veh_j$ do not enter into $\dz_{ij}$ even when both vehicles are applying avoidance maneuver and hence can apply \textit{any} control from each other's perspective. Thus at time $t$, $\veh_i$ not only needs to avoid the states that $\veh_j$ could be in at time $t$, but also all the states that could lead it to $\dz_{ij}$ \textit{in future} under some control actions of $\veh_i$ and $\veh_j$. To compute this set of states, we make the following key observation:
\begin{observation} \label{obs1_case1_caseB}
For computing $_1^B\ioset_i^j(t)$, it is sufficient to consider $\tsa_i = t$. If $\tsa_i > t$, then $\veh_i$ is not applying any avoidance maneuver at time $t$ and hence should only avoid the states that $\veh_j$ could be in at time $t$. However, this is already ensured during computation of $_1^A\ioset_i^j(t)$. If $\tsa_i < t$, then for a given $\tsa$, $\veh_i$ still needs to avoid the same set of states at time $t$ that it would have if $\tsa_i = t$.  
\end{observation}

Due to the separation and buffer regions, we have $\tsa_i - \tsa_j \geq \brd$. This along with Observation \ref{obs1_case1_caseB} implies that $\tsa_j \leq t - \brd$. Also, from Observation \ref{obs1_case1Obs}, we have $\tsa = \tsa_j \geq t-\iat$. Thus, $\tsa_j \in [t-\iat, t-\brd]$. Since the intruder is present for a maximum duration of $\iat$, $\veh_j$ might be applying any control during $[\tsa_j, \tsa_j + \iat]$ from the perspective of $\veh_i$. In particular, for any given $\tsa_j$, $\veh_j$ can reach any state in $\frs_{j}^{\mathcal{O}}(\tsa_j, t')$ at time $t' \in [\tsa_j, \tsa_j+\iat]$, starting from some state in $\boset_j(\tsa_j)$ at time $\tsa_j$. Here, $\frs_{j}^{\mathcal{O}}(\tsa_j, t')$ represents the FRS of $\boset_j(\tsa_j)$ computed forward for a duration of $(t' - \tsa_j)$ and is given by \eqref{eq:ObsFRS_case1_caseA}. 

Taking into account all possible $\tsa_j \in [t-\iat, t-\brd]$, $\state_j(\tau)$ is contained in the set:
\begin{equation}
\mathcal{K}^{\text{B1}}(\tau) = \bigcup_{\tsa_j \in [\tau-\iat, t-\brd]} \frs_{j}^{\mathcal{O}}(\tsa_j, \tau)
\end{equation} 
at time $\tau \in [t, t-\brd+\iat]$, where the upper bound on $\tau$ corresponds to the upper bound on $\tsa_j$. From Observation \ref{obs1_case1_caseA}, we have $\frs_{j}^{\mathcal{O}}(\tsa_j, \tau) \subseteq \frs_{j}^{\mathcal{O}}(\tau-\iat, \tau)$ for all $\tsa_j \in [\tau-\iat, t-\brd]$. Therefore, $\mathcal{K}^{\text{B1}}(\tau) = \frs_{j}^{\mathcal{O}}(\tau-\iat, \tau)$.

From the perspective of $\veh_i$, it needs to avoid all states at time $t$ that can reach $\mathcal{K}^{\text{B1}}(\tau)$ for some control action of $\veh_i$ during time duration $[t, \tau]$. This will ensure that $\veh_i$ and $\veh_j$ will not enter into each other's danger zones regardless of the avoidance maneuver applied by them. This set of states is given by the following BRS:
\begin{equation}  \label{eq:ObsBRS_case1_caseB}
\begin{aligned}
\brs^{\text{B1}}_{i}(t, t-\brd+\iat) = & \{y: \exists \ctrl_i(\cdot) \in \cfset_i, \exists \dstb_i(\cdot) \in \dfset_i, \\
& \state_i(\cdot) \text{ satisfies \eqref{eq:dyn}}, \state_i(t) = y, \\
& \exists s \in [t, t-\brd+\iat],\\
& \state_i(s) \in \tilde{\mathcal{K}}^{\text{B1}}(s)\},
\end{aligned}
\end{equation}
where
\begin{equation*}
\tilde{\mathcal{K}}^{\text{B1}}(s) = \{\state_j: \exists (y, h) \in \mathcal{K}^{\text{B1}}(s), \|\pos_j - y\|_2 \le \rc \}.
\end{equation*} 
The Hamiltonian $\ham^{\text{B1}}_{i}$ to compute $\brs^{\text{B1}}_{i}(\cdot)$ is given by
\begin{equation} \label{eqn:BRS_obsham_case1_caseB}
\ham^{\text{B1}}_{i}(\state_i, \costate) = \min_{\ctrl_i \in \cset_i, \dstb_i \in \dset_i} \costate \cdot f_i (\state_i, \ctrl_i, \dstb_i).
\end{equation}
Finally, the induced obstacle in this case is given by
\begin{equation} \label{eq:intruderObs_case1_caseB} 
_1^B\ioset_i^j(t) = \brs^{\text{B1}}_{i}(t, t-\brd+\iat).
\end{equation}
\end{itemize}
\subsection{Separation and Buffer Regions - Case 2} \label{sec:case2_maintext}
We now consider Case 2: $\tsa_i < \tsa_j, \tsa_i <\infty$. In this case, the intruder forces $\veh_i$, the lower-priority vehicle, to apply avoidance control before $\veh_j$, the higher-priority vehicle. 
The separation region, the buffer region and the obstacles in this case can be computed in a similar manner to that in Case 1.
The buffer region in this case is denoted as $\buff_{ji}(t)$ to differentiate it from Case 1 (i.e., the order of $i$ and $j$ indexes has been switched). 
Similarly, the obstacles for Case 2 are denoted as $_2^A\ioset_i^j(t)$ and $_2^B\ioset_i^j(t)$, corresponding to the two cases similar to that in Section \eqref{sec:intruderObs_case1}. 
For brevity purposes, this computation is presented in the Appendix.

\subsection{Trajectory Planning} \label{sec:path_planning}
In this section, our goal is to plan the trajectory of each vehicle such that it is guaranteed to safely reach its target in the absence of an intruder, and to ensure collision avoidance with vehicles or obstacles if forced to apply an avoidance maneuver. We also need to make sure that the trajectories of the vehicles are such that the separation requirement is satisfied at all times. To obtain such a trajectory, we take into account all the ``obstacles" computed in previous sections which ensure that the vehicle $\veh_i$ will not collide with any other vehicle, as long as it is outside these obstacles.   

Before, we plan such a trajectory, we need to compute one final set of obstacles. In particular, we need to compute the set of states states that $\veh_i$ needs to avoid in order to avoid a collision with static obstacles while it is applying an avoidance maneuver. Since $\veh_i$ applies avoidance maneuver for a maximum duration of $\iat$, this set is given by the following BRS:
\begin{equation} \label{eq:ObsBRS_static}
\begin{aligned}
\brs^{\text{S}}_{i}(t, t+\iat) = & \{y: \exists \ctrl_i(\cdot) \in \cfset_i, \exists \dstb_i(\cdot) \in \dfset_i, \\
& \state_i(\cdot) \text{ satisfies \eqref{eq:dyn}}, \state_i(t) = y, \\
& \exists s \in [t, t+\iat], \state_i(s) \in \mathcal{K}^{\text{S}}(s) \},\\
\mathcal{K}^{\text{S}}(s) = & \{\state_i: \exists (y, h) \in \soset_i, \|\pos_i - y\|_2 \le \rc \}.
\end{aligned}
\end{equation}
The Hamiltonian $\ham^{\text{S}}_{i}$ to compute $\brs^{\text{S}}_{i}(t, t+\iat)$ is given by:
\begin{equation} \label{eqn:BRS_obsham_static}
\ham^{\text{S}}_{i}(\state_i, \costate) = \min_{\ctrl_i \in \cset_i, \dstb_i \in \dset_i} \costate \cdot f_i (\state_i, \ctrl_i, \dstb_i).
\end{equation}
$\brs^{\text{S}}_{i}(t, t+\iat)$ represents the set of all states of $\veh_i$ at time $t$ that can lead to a collision with a static obstacle for some time $\tau > t$ for some control strategy of $\veh_i$.

During the trajectory planning of $\veh_i$, if we use $\buff_{ij}(t)$ and $\buff_{ji}(t)$ as obstacles at time $t$, then the separation requirement is ensured between $\veh_i$ and $\veh_j$ for all intruder strategies and $\tsa = t$. Similarly, if obstacles computed in sections \ref{sec:intruderObs_case1} and \ref{sec:case2_maintext} are used as obstacles in trajectory planning, then we can guarantee collision avoidance between $\veh_i$ and $\veh_j$ while they are avoiding the intruder. Thus, the overall obstacle for $\veh_i$ is given by:
\begin{equation} \label{eq:obsseti_intr}
\begin{aligned}
\obsset_i(t)  =  & \brs^{\text{S}}_{i}(t, t+\iat) \bigcup \\
& \bigcup_{j=1}^{i-1} \left( \buff_{ij}(t) \cup \buff_{ji}(t) \bigcup_{k \in\{1, 2\}} {}_k^A\ioset_i^j(t) \bigcup_{k \in\{1, 2\}} {}_k^B\ioset_i^j(t) \right).
\end{aligned}
\end{equation}

Given $\obsset_i(t)$, we compute a BRS $\brs^{\text{AO}}_{i}(t, \sta_i)$ for trajectory planning that contains the initial state of $\veh_i$ and avoids these obstacles:
\begin{equation} \label{eqn:intrBRS1}
\begin{aligned}
\brs^{\text{PP}}_{i}(t, \sta_i) = & \{y: \exists \ctrl_i(\cdot) \in \cfset_i, \forall \dstb_i(\cdot) \in \dfset_i, \\
& \state_i(\cdot) \text{ satisfies \eqref{eq:dyn}}, \forall s \in [t, \sta_i], \state_i(s) \notin \obsset_i(s), \\
& \exists s \in [t, \sta_i], \state_i(s) \in \targetset_i, \state_i(t) = y \}.
\end{aligned}
\end{equation}
The Hamiltonian $\ham^{\text{PP}}_{i}$ to compute the BRS in \eqref{eqn:intrBRS1} is given by:
\begin{equation} \label{eqn:BRSham}
\ham^{\text{PP}}_{i}(\state_i, \costate) = \min_{\ctrl_i \in \cset_i} \max_{\dstb_i \in \dset_i} \costate \cdot f_i (\state_i, \ctrl_i, \dstb_i)
\end{equation}

Note that $\brs^{\text{PP}}_{i}(\cdot)$ ensures goal satisfaction for $\veh_i$ in the absence of intruder. The goal satisfaction controller is given by:
\begin{equation} \label{eqn:PPPolicy}
{\ctrl^{\text{PP}}_{i}}(t, \state_i) = \arg \min_{\ctrl_i \in \cset_i} \max_{\dstb_i \in \dset_i} \costate \cdot f_i (\state_i, \ctrl_i, \dstb_i)
\end{equation}
When intruder is not present in the system, $\veh_i$ applies the control ${\ctrl^{\text{PP}}_{i}}$ and we get the ``nominal trajectory" of $\veh_i$. Once intruder appears in the system, $\veh_i$ applies the avoidance control ${\ctrl^{\text{A}}_i}$ and hence might deviate from its nominal trajectory. The overall control policy for avoiding the intruder and collision with other vehicles is thus given by:
\begin{equation} \label{eqn:final_nominalcontroller}
{\ctrl^*_{i}}(t) = 
\left \{ 
\begin{array}{ll}
{\ctrl^{\text{PP}}_{i}}(t) & t \leq \tsa_i\\
{\ctrl^{\text{A}}_{i}}(t) & \tsa_i \leq t \leq \tsa + \iat
\end{array}
\right.
\end{equation}

If $\veh_i$ starts within $\brs^{\text{PP}}_{i}$ and uses the control ${\ctrl^*_{i}}$, it is guaranteed to avoid collision with the intruder and other STP vehicles, regardless of the control strategy of $\veh_{\intr}$. Finally, since we use separation and buffer regions as obstacles during the trajectory planning of $\veh_i$, it is guaranteed that $|\tsa_i - \tsa_j| \geq \brd$ for all $j < i$. Therefore, atmost $\nva$ vehicles are forced to apply an avoidance maneuver. The planning phase is summarized in Algorithm \ref{alg:intruder_plan}.
\begin{remark}
Note that given $\brs^{\text{PP}}_{i}$ and ${\ctrl^{\text{PP}}_{i}}$, the base obstacles required for the computation of the separation region in Section \ref{sec:sepRegion_case1} can be computed using equations (25), (31) and (37) in \cite{Chen2016d}. Also, if we use the robust trajectory tracking method to compute the base obstacles, we would need to augment the obstacles in \eqref{eq:obsseti_intr} by the error bound of $\veh_i$, $\disckernel_i$ (for details, see Section 5A-3 in \cite{Chen2016d}). The BRS in \eqref{eqn:intrBRS1} in this case is computed assuming no disturbance in $\veh_i$'s dynamics.
\end{remark}
\begin{algorithm}[tb!]
\SetKwInOut{Input}{input}
\SetKwInOut{Output}{output}
	\DontPrintSemicolon
	\caption{The intruder avoidance algorithm: Planning-phase (offline planning)}
	\label{alg:intruder_plan}
	\Input{Set of vehicles $\veh_i$ in the descending priority order, their dynamics \eqref{eq:dyn} and initial states $\state_i^0$;\newline
	Vehicle destinations $\targetset_i$ and static obstacles $\soset_i$;\newline
	Intruder dynamics $f_{\intr}$ and the maximum avoidance time $\iat$ ;\newline
	Maximum number of vehicles allowed to re-plan their trajectories $\nva$.}
    \Output{The nominal controller $\ctrl^{\text{PP}}$ and the avoidance controller $\ctrl^{\text{A}}$ for all vehicles.}
	\For{\text{$i=1:N$}}{
			\textbf{Avoid region and avoidance control for $\veh_{i}$} \;	
			compute the avoid region $\brs^{\text{A}}_{i}$ using \eqref{eqn:avoidBRS}; \;
			compute the avoidance controller $\ctrl^{\text{A}}_i$ using \eqref{eqn:optAvoidCtrl}; \;
			output the optimal avoidance controller $\ctrl_i^{\text{A}}$ for $\veh_i$. \;
			\If{\text{$i \neq 1$}}{
			    \textbf{Computation of separation region for $\veh_{i}$} \;	
				\For{\text{$j=1:i-1$}}{
					given the base obstacles $\boset_j(\cdot)$ and the avoid region $\brs^{\text{A}}_{j}$, compute the separation regions in \eqref{eqn:sepRegion_case1} and \eqref{eqn:sepRegion_case2}; 					\;}
				\textbf{Computation of buffer region for $\veh_{i}$} \;	
				\For{\text{$j=1:i-1$}}{
					given the separation regions, compute the relative buffer regions $\brs^{\text{B}}$ in \eqref{eqn:buffBRS_case1} and \eqref{eqn:buffBRS_case2}; \;
					given the relative buffer regions, compute the buffer regions in \eqref{eqn:buffRegion_case1} and \eqref{eqn:buffRegion_case2}; \;}
			}
			\textbf{Computation of obstacles for $\veh_{i}$} \;
			\If{\text{$i \neq 1$}}{
				\For{\text{$j=1:i-1$}}{
					given the base obstacles $\boset_j(\cdot)$, compute the obstacles $_1^A\ioset_i^j(t)$ in \eqref{eq:intruderObs_case1_caseA}, $_1^B\ioset_i^j(t)$ in \eqref{eq:intruderObs_case1_caseB}, $_2^A\ioset_i^j(t)$ in \eqref{eq:intruderObs_case2_caseA}, and $_2^B\ioset_i^j(t)$ in \eqref{eq:intruderObs_case2_caseB}; \;}
			}
			compute the effective static obstacle to avoid ($\brs^{\text{S}}_{i}$) using \eqref{eq:ObsBRS_static}; \;
			\textbf{Trajectory planning for $\veh_{i}$} \;
			compute the total obstacle set $\obsset_i(t)$ given by \eqref{eq:obsseti_intr};\;
			compute the BRS $\brs^{\text{PP}}_{i}(t, \sta_i)$ defined in \eqref{eqn:intrBRS1};\;
			\textbf{The nominal controller of $\veh_{i}$} \;
			compute the nominal controller ${\ctrl^{\text{PP}}_{i}}(\cdot)$ given by \eqref{eqn:PPPolicy};\;
			output the nominal controller for $\veh_i$.\;
			\textbf{Base obstacle induced by $\veh_{i}$} \;
			given the nominal controller ${\ctrl^{\text{PP}}_{i}}(\cdot)$ and the BRS $\brs^{\text{PP}}_{i}(t, \sta_i)$, compute the base obstacles $\boset_i(\cdot)$ using equations (25), (31) or (37) in \cite{Chen2016d}, depending on the information assumed to be known about the higher-priority vehicles.
		}
\end{algorithm}
%
\subsection{Replanning after intruder avoidance} \label{sec:replan}
As discussed in Section \ref{sec:path_planning}, the intruder can force some STP vehicles to deviate from their planned nominal trajectory; therefore, goal satisfaction is no longer guaranteed once a vehicle is forced to apply an avoidance maneuver. Therefore, we have to replan the trajectories of these vehicles once $\veh_{\intr}$ disappears. The set of all vehicles $\veh_i$ for whom replanning is required, $\rvs$, can be obtained by checking if a vehicle $\veh_i$ applied any avoidance control during $[\tsa, \tsa+\iat]$, i.e.,
\begin{equation} \label{eq:RVS}
\rvs = \{\veh_i: \tsa_i < \infty, i \in \{1, \ldots, \N \} \}. 
\end{equation}  

Note that due to the presence of separation and buffer regions, at most $\nva$ vehicles can be affected by $\veh_{\intr}$, i.e., $|\rvs| \leq \nva$. Goal satisfaction controllers which ensure that these vehicles reach their destinations can be obtained by solving a new STP problem, where the starting states of the vehicles are now given by the states they end up in, denoted $\tilde{\state}_i^0$, after avoiding the intruder. Note that we can pick $\nva$ beforehand and design buffer regions accordingly. Thus, by picking compatible $\nva$ based on the available computation resources during run-time, we can ensure that this replanning can be done in real time. Moreover, flexible trajectory-planning algorithms such as FaSTrack \cite{Herbert2017} can be used that can perform replanning efficiently in real-time.   

Let the optimal control policy corresponding to this liveness controller be denoted ${\ctrl^{\text{L}}_{i}}(t, \state_i)$. The overall control policy that ensures intruder avoidance, collision avoidance with other vehicles, and successful transition to the destination for vehicles in $\rvs$ is given by:

\begin{equation} \label{eqn:full_controller}
\ctrl_i^{\text{RP}}(t) = 
\left \{ 
\begin{array}{ll}
{\ctrl^*_{i}}(t, \state_i) & t \leq \tsa + \iat\\
{\ctrl^{\text{L}}_{i}}(t, \state_i) & t > \tsa + \iat
\end{array}
\right.
\end{equation}

Note that in order to re-plan using a STP method, we need to determine feasible $\sta_i$ for all vehicles. This can be done by computing an FRS:
\begin{equation} \label{eq:re-planFRS}
\begin{aligned} 
\frs_i^{\text{RP}}(\tea, t) = & \{y \in \R^{n_i}: \exists \ctrl_i(\cdot) \in \cfset_i, \forall \dstb_i(\cdot) \in \dfset_i, \\
& \state_i(\cdot) \text{ satisfies \eqref{eq:dyn}}, \state_i(\tea) = \tilde{\state}_i^0, \\
& \state_i(t) = y, \forall s \in [\tea, t], \state_i(s) \notin \obsset_i^{\text{RP}}(s) \},
\end{aligned}
\end{equation}
\noindent where $\tilde{\state}_i^0$ represents the state of $\veh_i$ at $t = \tsa+\iat$; $\obsset_i^{\text{RP}}(\cdot)$ takes into account the fact that $\veh_i$ now needs to avoid all other vehicles in $(\rvs)^C$ and is defined in a way analogous to \eqref{eq:obsseti}. The FRS in \eqref{eq:re-planFRS} can be obtained by solving 
\begin{equation}
\begin{aligned}
\max \Big\{&D_t \valfuncfwd_i^{\text{RP}}(t, \state_i) + \ham_i^{\text{RP}}(t, \state_i, \nabla \valfuncfwd_i^{\text{RP}}(t, \state_i)), \\
&\qquad - \obsfunc_i^{\text{RP}}(t, \state_i) - \valfuncfwd_i^{\text{RP}}(t, \state_i) \Big\} = 0\\
&\valfuncfwd_i^{\text{RP}}(\tsa, \state_i) = \max\{\fc_i^{\text{RP}}(\state_i), -\obsfunc_i^{\text{RP}}(\tsa, \state_i)\} \\
&\ham_i^{\text{RP}}(\state_i, \costate) = \max_{\ctrl_i \in \cset_i} \min_{\dstb_i \in \dset_i} \costate \cdot f_i (\state_i, \ctrl_i, \dstb_i)
\end{aligned}
\end{equation} 

\noindent where $\valfuncfwd_i^{\text{RP}}, \obsfunc_i^{\text{RP}}, \fc_i^{\text{RP}}$ represent the FRS, obstacles during re-planning, and the initial state of $\veh_i$, respectively. The new $\sta_i$ of $\veh_i$ is now given by the earliest time at which $\frs_i^{\text{RP}}(\tea, t)$ intersects the target set $\targetset_i$, $\sta_i := \arg \inf_t \{ \frs_i^{\text{RP}}(\tea, t) \cap \targetset_i \neq \emptyset \}$. Intuitively, this means that there exists a control policy which will steer the vehicle $\veh_i$ to its destination by that time, despite the worst case disturbance it might experience. The replanning phase is summarized in Algorithm \ref{alg:intruder_replan}.
\begin{remark}
Note that even though we have presented the analysis for one intruder, the proposed method can handle multiple intruders as long as only one intruder is present \textit{at any given time}.
\end{remark}
\begin{algorithm}
\SetKwInOut{Input}{input}
\SetKwInOut{Output}{output}
	\DontPrintSemicolon
	\caption{The intruder avoidance algorithm: Replanning-phase (real-time planning)}
	\label{alg:intruder_replan}
	\Input{Set of vehicles $\veh_i \in \rvs$ that require replanning;\newline
	Vehicle dynamics \eqref{eq:dyn} and new initial states $\tilde{\state}_i^0$;\newline
	Vehicle destinations $\targetset_i$ and static obstacles $\soset_i$;\newline
	Total base obstacle set $\obsset_i^{\text{RP}}(\cdot)$ induced by all other vehicles in $(\rvs)^C$.}
    \Output{The updated nominal controller ${\ctrl^{\text{L}}_{i}}$ for all vehicles in $\rvs$.}
	\For{\text{$\veh_i \in \rvs$}}{
			\textbf{Computation of the updated $\sta_i$ for $\veh_{i}$} \;	
			given the obstacle set $\obsset_i^{\text{RP}}(\cdot)$, compute the FRS $\frs_i^{\text{RP}}(\tea, t)$ in \eqref{eq:re-planFRS}; \;
			the updated $\sta_i$ for $\veh_i$ is given by $\arg \inf_t \{ \frs_i^{\text{RP}}(\tea, t) \cap \targetset_i \neq \emptyset \}$. \;
			\textbf{Trajectory and controller of $\veh_{i}$} \;
			given the updated STA $\sta_i$, the initial state $\tilde{\state}_i^0$, the total obstacle set $\obsset_i^{\text{RP}}(\cdot)$, the vehicle dynamics \eqref{eq:dyn} and the target set $\targetset_i$, use Algorithm \ref{alg:intruder_plan} to replan the nominal trajectory and controller.
		}
\end{algorithm}
\section{Simulations \label{sec:simulations}}
We now illustrate the proposed algorithm using a fifty-vehicle example. 

\subsection{Setup \label{sec:simSetup}}
\begin{figure}
  \centering
  \includegraphics[width=0.8\columnwidth]{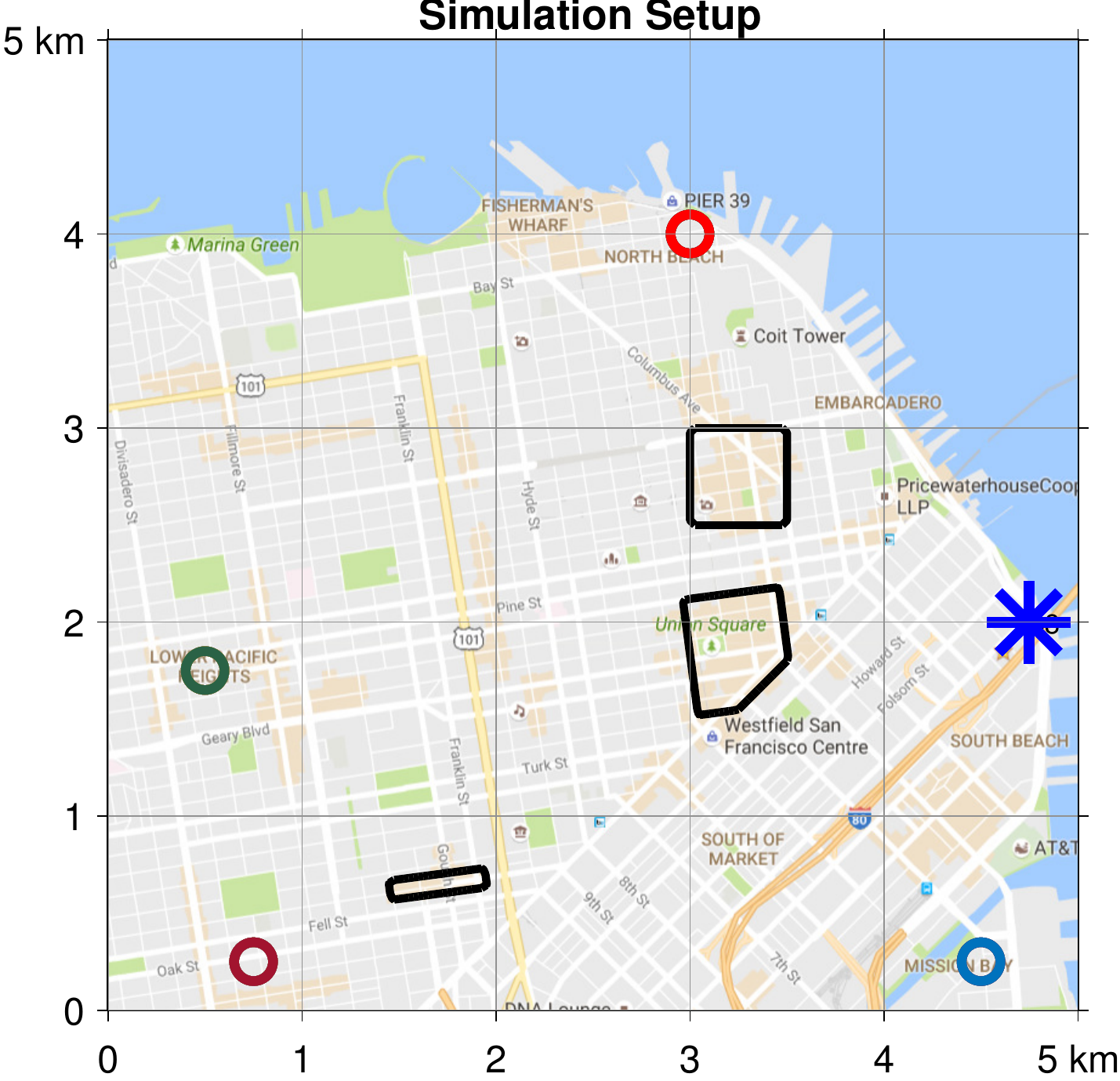}
  \caption{Simulation setup. A \SI{25}{\km\squared} area of San Francisco city is used as the state-space for vehicles. STP vehicles originate from the Blue star and go to one of the four destinations, denoted by the circles. Tall buildings in the downtown area are used as static obstacles, represented by the black contours.}
  \label{fig:sf_setup}
\end{figure}

Our goal is to simulate a scenario where UAVs are flying through an urban environment. This setup can be representative of many UAV applications, such as package delivery, aerial surveillance, etc. For this purpose, we use the city of San Francisco (SF), California, USA as our planning region, as shown in Figure \ref{fig:sf_setup}. Practically speaking, depending on the UAV density, it may be desirable to have smaller planning regions that together cover the SF area; however, such considerations are out of the scope of this paper.

Each box in Figure \ref{fig:sf_setup} represents a \SI{500}{\m} $\times$ \SI{500}{\m} area of SF. The origin point for the vehicles is denoted by the Blue star. Four different areas in the city are chosen as the destinations for the vehicles. Mathematically, the target sets $\targetset_i$ of the vehicles are circles of radius $r$ in the position space, i.e. each vehicle is trying to reach some desired set of positions. In terms of the state space $\state_i$, the target sets are defined as $\targetset_i = \{\state_i: \|\pos_i - c_i\|_2 \le 100 \text{m}\}$, where $c_i$ are centers of the target circles. The four targets are represented by four circles in Figure \ref{fig:sf_setup}. The destination of each vehicle is chosen randomly from these four destinations. Finally, tall buildings in downtown San Francisco are used as static obstacles, denoted by black contours in Figure \ref{fig:sf_setup}. We use the following dynamics for each vehicle:
\begin{equation}
\label{eq:dyn_i}
\begin{aligned}
\dot{\pos}_{x,i} &= v_i \cos \theta_i + d_{x,i} \\
\dot{\pos}_{y,i} &= v_i \sin \theta_i + d_{y,i}\\
\dot{\theta}_i &= \omega_i, \\
\underline{v} \le v_i \le \bar{v}, & ~|\omega_i| \le \bar{\omega}, ~\|(d_{x,i}, d_{y,i}) \|_2 \le d_{r},
\end{aligned}
\end{equation}
\noindent where $\state_i = (\pos_{x,i}, \pos_{y,i}, \theta_i)$ is the state of vehicle $\veh_i$, $\pos_i = (\pos_{x,i}, \pos_{y,i})$ is the position and $\theta_i$ is the heading. $d = (d_{x,i}, d_{y,i})$ represents $\veh_i$'s disturbances, for example wind, that affect its position evolution. The control of $\veh_i$ is $u_i = (v_i, \omega_i)$, where $v_i$ is the speed of $\veh_i$ and $\omega_i$ is the turn rate; both controls have a lower and upper bound. To make our simulations as close as possible to real scenarios, we choose velocity and turn-rate bounds as $\underline{v} = $\SI{0}{\m\per\s}, $\bar{v} = $\SI{25}{\m\per\s}, $\bar\omega = $\SI{2}{\radian\per\s}, aligned with the modern UAV specifications \cite{UAVspecs1, UAVspecs2}. The disturbance bound is chosen as $d_{r} = $\SI{6}{\m\per\s}, which corresponds to \textit{moderate winds} on the Beaufort wind force scale \cite{Windscale}. Note that we have used same dynamics and input bounds across all vehicles for clarity of illustration; however, our method can easily handle more general systems of the form in which the vehicles have different control bounds and dynamics.

The goal of the vehicles is to reach their destinations while avoiding a collision with the other vehicles or the static obstacles. The vehicles also need to account for the possibility of the presence of an intruder for a maximum duration of $\iat = $\SI{10}{\s}, whose dynamics are given by \eqref{eq:dyn_i}. The joint state space of this fifty-vehicle system is 150-dimensional (150D); therefore, we assign a priority order to vehicles and solve the trajectory planning problem sequentially. For this simulation, we assign a random priority order to fifty vehicles and use the algorithm proposed in Section \ref{sec:intruder} to compute a separation between STP vehicles so that they do not collide with each other or the intruder. 

\subsection{Results \label{sec:simResults}}
\begin{figure}
  \centering
  \includegraphics[width=0.7\columnwidth]{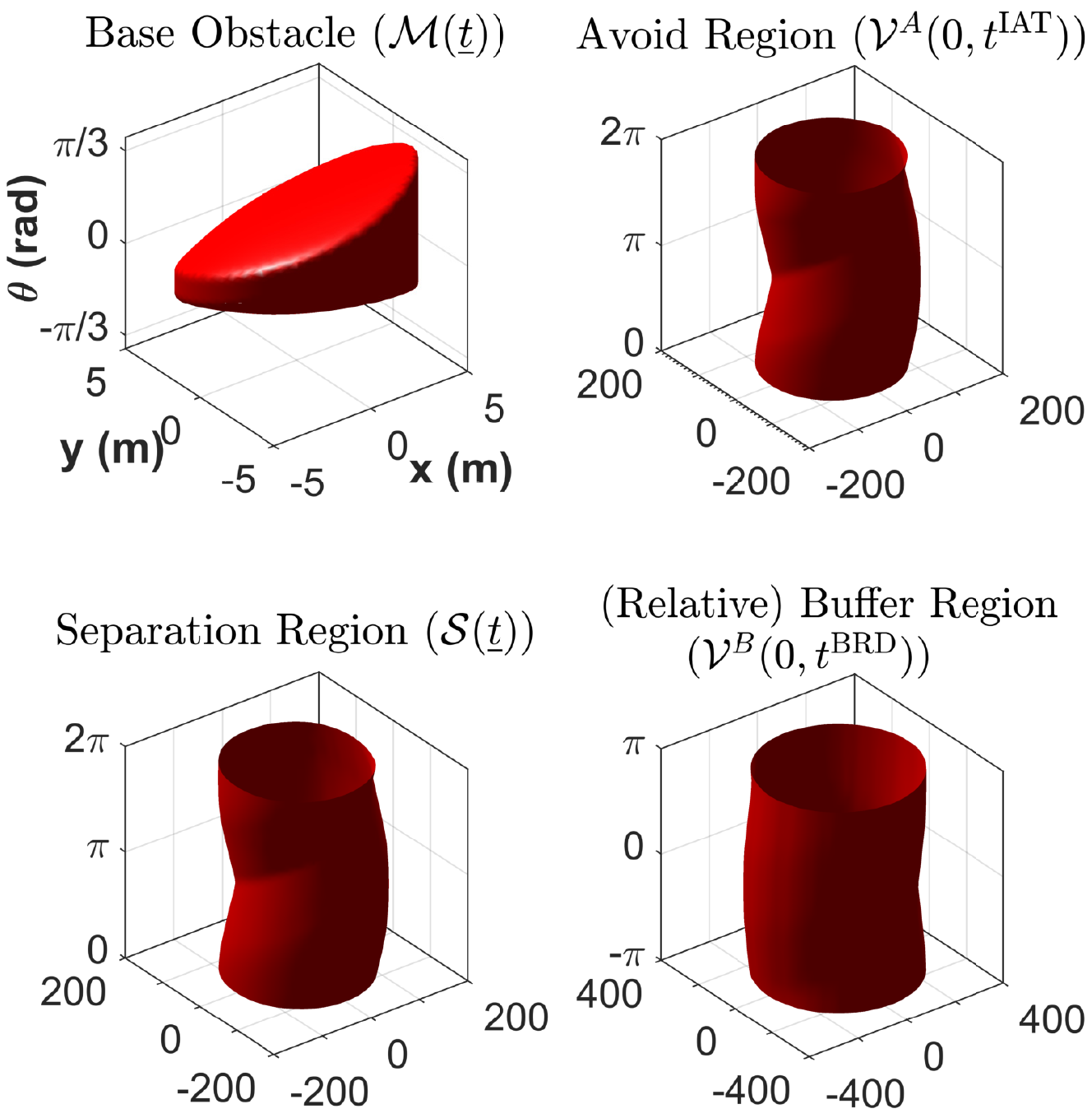}
  \caption{Base obstacle $\boset(t)$ , Avoid region $\brs^{\text{A}}(0, \iat)$, Separation region $\sep(t)$ and Relative buffer region $\brs^{\text{B}}(0, \brd)$ for vehicles. The three axes represent three states of the vehicles.}
  \label{fig:MaxMin}
\end{figure}

In this section, we present the simulation results for $\nva = 3$; occasionally, we also compare the results for different values of $\nva$ to highlight some key insights about the proposed algorithm. As per Algorithm \ref{alg:intruder_plan}, we begin with computing the avoid region $\brs^{\text{A}}_{i}(0, \iat)$. To compute the avoid region, relative dynamics between $\veh_i$ and $\veh_{\intr}$ are required. Given the dynamics in \eqref{eq:dyn_i}, the relative dynamics are given by \cite{Mitchell05}:
\begin{equation}
\label{eq:reldyn_i}
\begin{aligned}
\dot{\pos}_{x, \intr, i} &= v_{\intr} \cos \theta_{\intr, i} - v_i + \omega_i {\pos}_{y, \intr, i} + d_{x,i} + d_{x,\intr}\\
\dot{\pos}_{y, \intr, i} &= v_i \sin \theta_{\intr, i} - \omega_i {\pos}_{x, \intr, i} + d_{y,i} + d_{y,\intr}\\
\dot{\theta}_{\intr, i} &= \omega_{\intr} - \omega_i,
\end{aligned}
\end{equation}    
where $\state_{\intr, i} = (\pos_{x, \intr, i}, \pos_{y, \intr, i}, \theta_{\intr, i})$ is the relative state between $\veh_{\intr}$ and $\veh_i$. Given the relative dynamics, the avoid region can be computed using \eqref{eqn:avoidBRS}. For all the BRS and FRS computations in this simulation, we use Level Set Toolbox \cite{Mitchell07b}. Also, since the vehicle dynamics are same across all vehicles, we will omit the vehicle index from sets wherever applicable. The avoid region $\brs^{\text{A}}(0, \iat)$ for STP vehicles is shown in the top-right plot of Figure \ref{fig:MaxMin}.

As long as $\veh_{\intr}$ starts outside the avoid region, $\veh_i$ is guaranteed to be able to avoid the intruder for a duration of $\iat$. Given $\brs^{\text{A}}(0, \iat)$, we can compute the minimum required detection range $\dsen$ given by \eqref{eqn:sen_distance} for the circular sensors, which turns out to be \SI{100}{\m} in this case, corresponding to a detection of \SI{4}{\s} in advance (given the speed of \SI{25}{\m\per\s}). So as long as the vehicles can detect the intruder within \SI{100}{\m}, the proposed algorithm guarantees collision avoidance with the intruder as well as a safe transit to their respective destinations.   

Next, we compute the separation and buffer regions between vehicles. For the computation of base obstacles, we use RTT method \cite{Bansal2017}. In RTT method, a nominal trajectory is declared by the higher-priority vehicles, which is then guaranteed to be tracked with some known error bound in the presence of disturbances. The base obstacles are thus given by a ``bubble" around the nominal trajectory. For further details of RTT method, we refer the interested readers to Section 4C in \cite{Bansal2017}. In presence of moderate winds, the obtained error bound is \SI{5}{\m}. This means that given any trajectory of vehicle, winds can at most cause a deviation of \SI{5}{\m} from this trajectory. The overall base obstacle $\boset$ around the point $(0, 0, 0)$ is shown in the top-left plot of Figure \ref{fig:MaxMin}. The base obstacles induced by a higher-priority vehicle are thus given by this set augmented on the nominal trajectory, the trajectory that a vehicle will follow if the intruder never appears in the system, and is obtained by executing the control policy ${\ctrl^{\text{PP}}}(\cdot)$ in \eqref{eqn:PPPolicy} for the higher-priority vehicles.

\begin{figure}
  \centering
  \includegraphics[width=0.6\columnwidth]{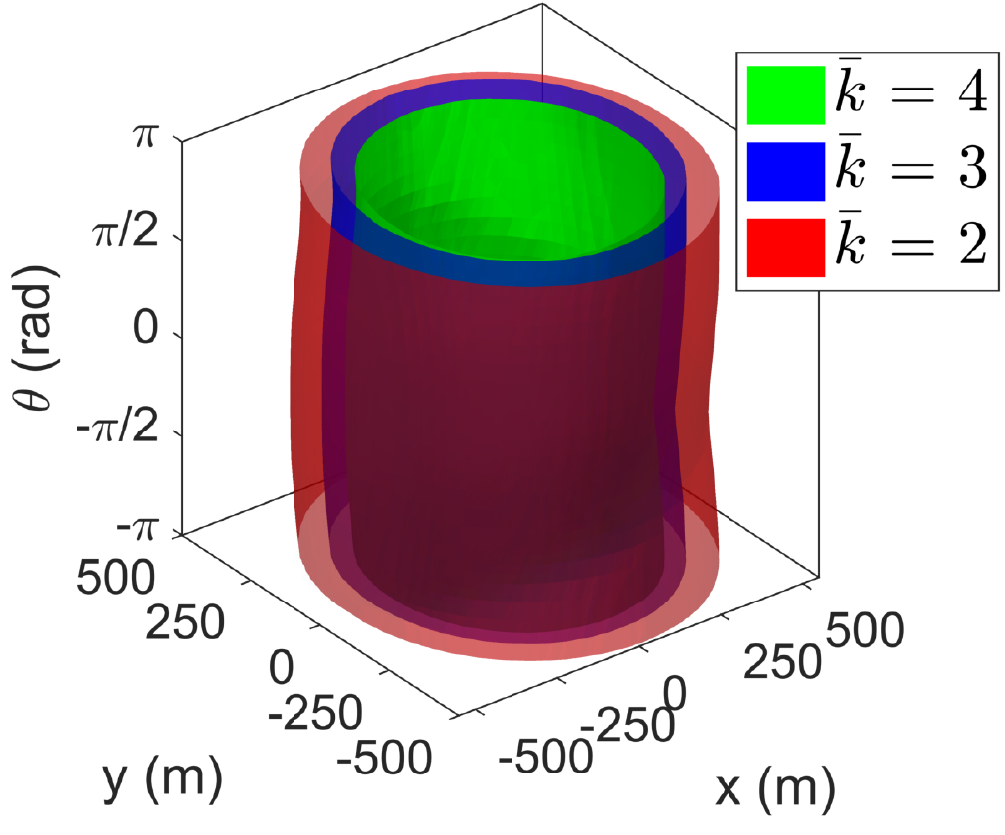}
  \caption{Buffer regions for different $\nva$ (best visualized with colors). As $\nva$ decreases, a larger buffer is required between vehicles to ensure that the intruder spends more time traveling through this buffer region so that it forces fewer vehicles to apply an avoidance maneuver.}
  \label{fig:buffRegions}
\end{figure}

Given $\boset$ of the higher-priority vehicles and $\brs^{\text{A}}(0, \iat)$, we compute the separation region $\sep$ as defined in \eqref{eqn:sepRegion_case1}. Relative buffer region $\brs^{\text{B}}(0, \brd)$, defined in \eqref{eqn:buffBRS_case1}, is similarly computed. The results are shown in the bottom two plots of Figure \ref{fig:MaxMin}. Finally, we compute the buffer region as defined in \eqref{eqn:buffRegion_case1}. The resultant buffer region is shown in Blue in Figure \ref{fig:buffRegions}. If $\veh_j$ is inside the base obstacle set and $\veh_i$ is outside the buffer region, we can ensure that the intruder will have to spend a duration of at least $\brd = $\SI{10/3}{\s} to go from the boundary of the avoid region of $\veh_j$ to the boundary of the avoid region of $\veh_i$. 

For the comparison purposes, we also computed the buffer regions for $\nva = 2$ and $\nva = 4$. 
As shown in Figure \ref{fig:buffRegions}, a bigger buffer is required between vehicles when $\nva$ is smaller. Intuitively, when $\nva$ is smaller, a larger buffer is required to ensure that the intruder spends more time ``traveling" through this buffer region so that it can affect fewer vehicles in the same duration.             

\begin{figure}
  \centering
  \includegraphics[width=0.8\columnwidth]{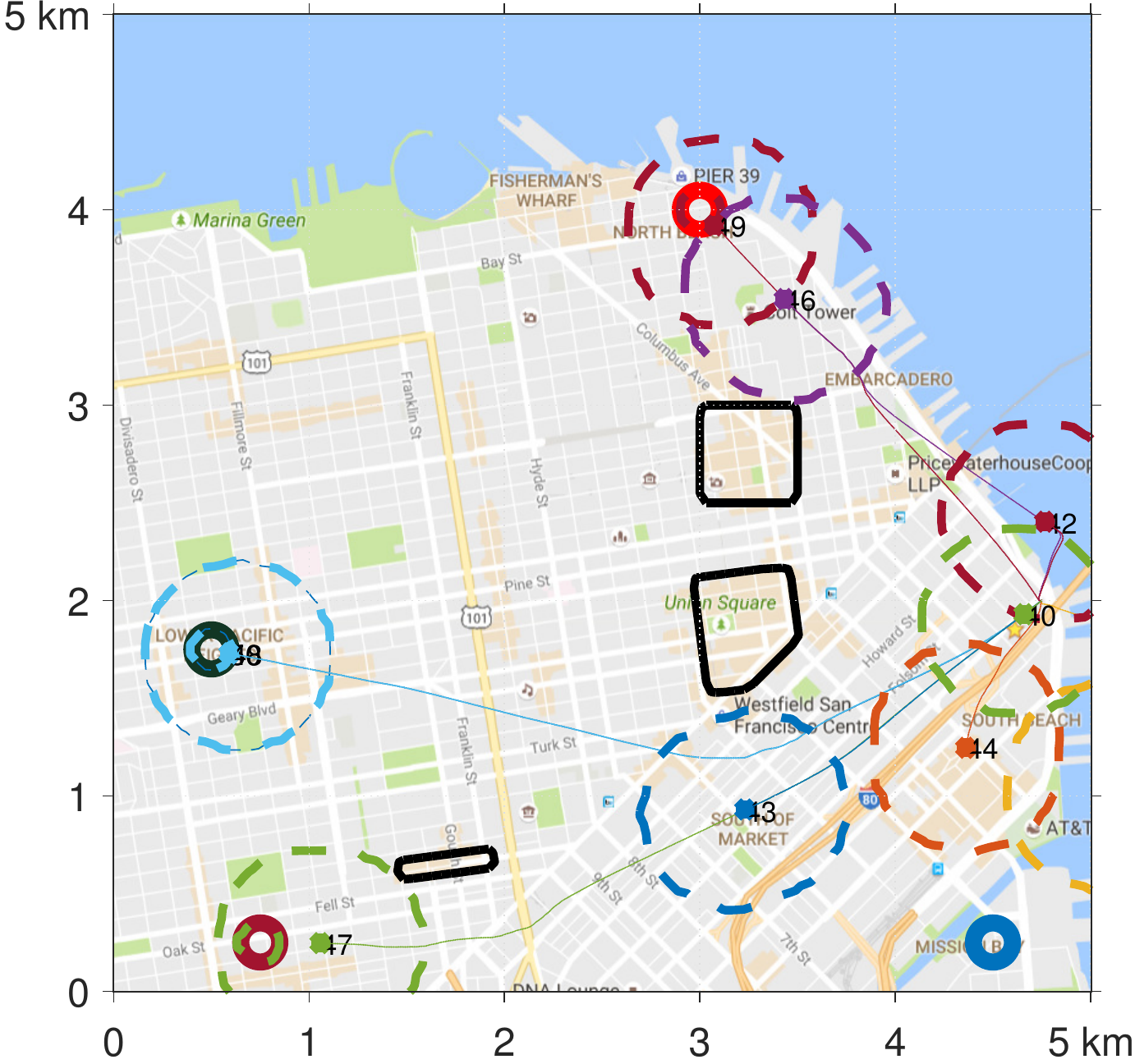}
  \caption{Nominal trajectories and induced obstacles by different vehicles. The nominal trajectories (solid lines) are well separated from each other to ensure that the intruder cannot force more than 3 vehicles to apply an avoidance maneuver.}
  \label{fig:trajObsSim}
\end{figure}

These buffer region computations along with the induced obstacle computations were similarly performed sequentially for each vehicle to obtain $\obsset(\cdot)$ in \eqref{eq:obsseti_intr}. This overall obstacle set was then used during their trajectory planning and the control policy $\ctrl^{\text{PP}}(\cdot)$ was computed, as defined in \eqref{eqn:PPPolicy}. Finally, the corresponding nominal trajectories were obtained by executing control policy $\ctrl^{\text{PP}}(\cdot)$. 
The nominal trajectories and the overall obstacles for different vehicles are shown in Figure \ref{fig:trajObsSim}. The numbers in the figure represent the vehicle numbers. The nominal trajectories (solid lines) are well separated from each other to ensure collision avoidance even during a worst-case intruder ``attack". At any given time, the vehicle density is low to ensure that the intruder cannot force more than three vehicles to apply an avoidance maneuver. This is also evident from large obstacles induced by vehicles for the lower priority vehicles (dashed circles). This lower density of vehicles is the price that we pay for ensuring that the replanning can be done efficiently in real-time. We discuss this trade-off further in section \ref{sec:discuss}.

\begin{figure}
  \centering
  \includegraphics[width=0.6\columnwidth]{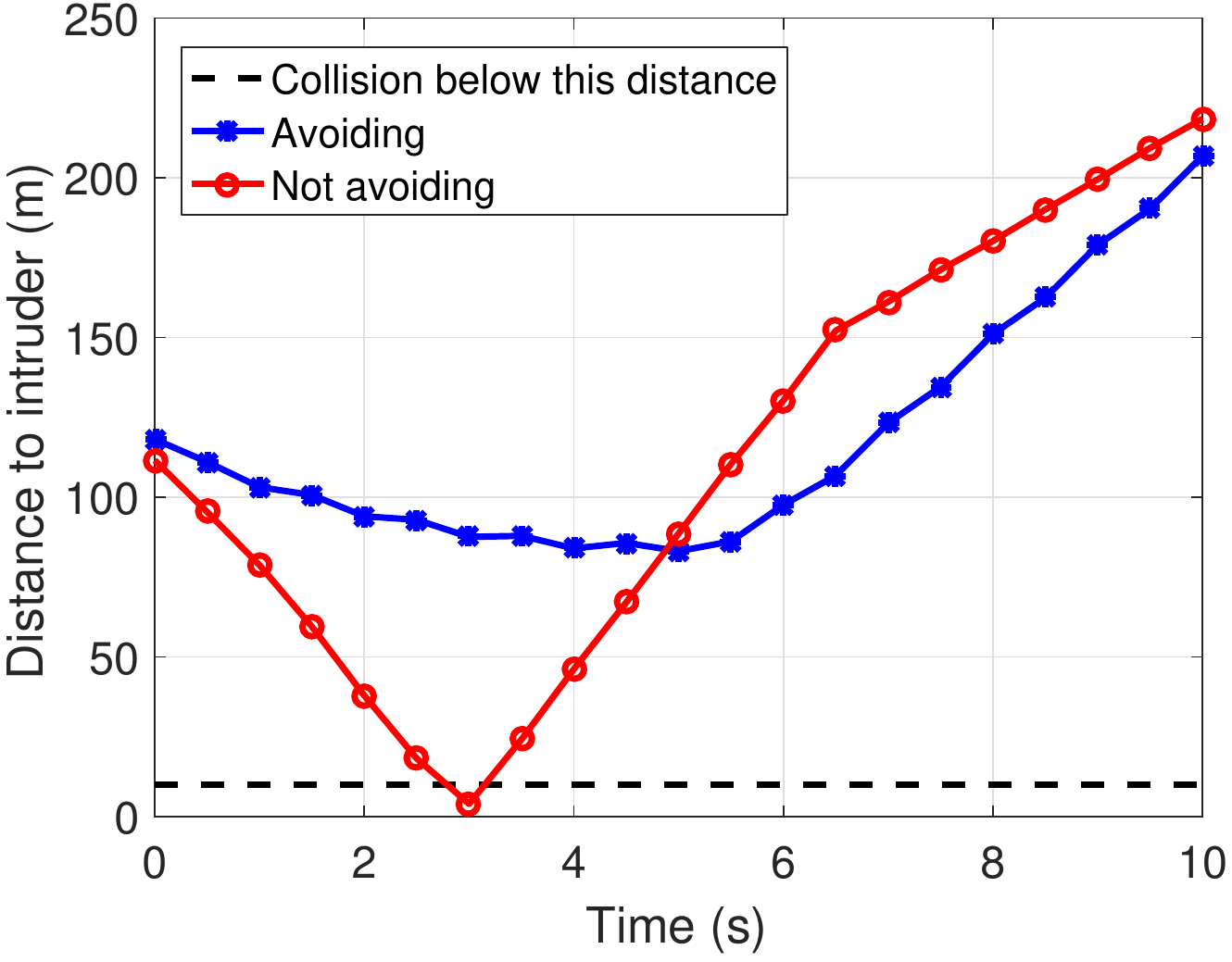}
  \caption{The trajectory of a STP vehicle when it applies the nominal controller vs when it applies the avoidance control. The vehicle is forced to apply the avoidance maneuver in the presence of an intruder, which can cause vehicle's deviation from its nominal trajectory.}
  \label{fig:trajComparison}
\end{figure}

In the absence of an intruder the vehicles transit successfully to their destinations with control policy $\ctrl^{\text{PP}}(\cdot)$, but they can deviate from the shown nominal trajectories if an intruder appears in the system. In particular, if a vehicle continues to apply the control policy $\ctrl^{\text{PP}}(\cdot)$ in the presence of an intruder, it might lead to a collision.
In Figure \ref{fig:trajComparison}, we plot the distance between an STP vehicle and the intruder when the vehicle applies the control policy $\ctrl^{\text{PP}}(\cdot)$ (Red line) vs when it applies ${\ctrl^{\text{A}}}$ (Blue line). Black dashed line represents the collision radius $r=$\SI{100}{\m} between the vehicle and the intruder. As evident from the figure, if the vehicle continues to apply the control policy $\ctrl^{\text{PP}}(\cdot)$ in the presence of an intruder, the intruder enters in its danger zone. Thus, it is forced to apply the avoidance control, which can cause a deviation from the nominal trajectory, but will successfully avoid the intruder.

\begin{figure}
\begin{subfigure}{.5\columnwidth}
  \centering
  \includegraphics[width=\columnwidth]{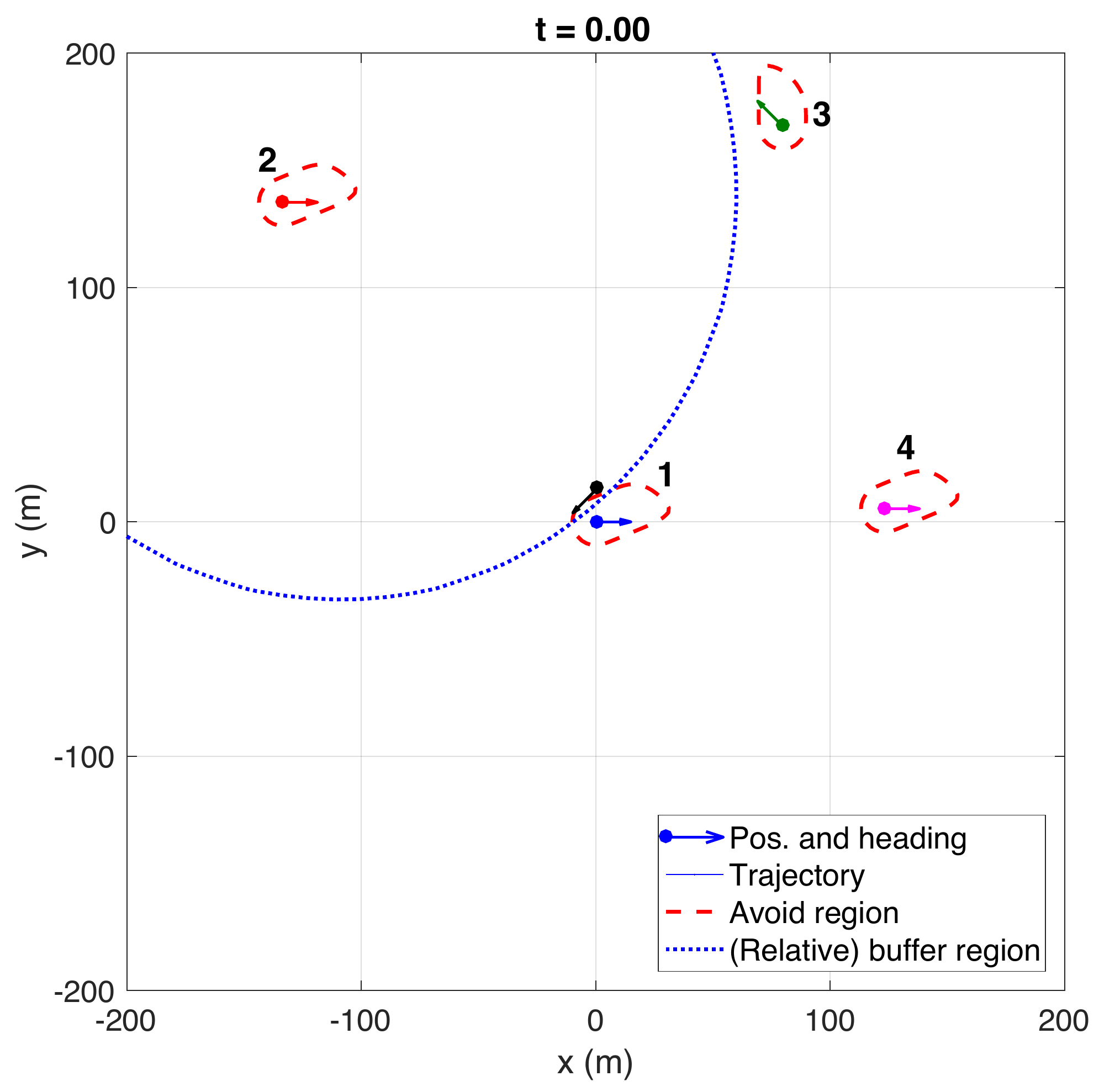}
  \subcaption{}
  \label{fig:worstcase1}
\end{subfigure}%
\begin{subfigure}{.5\columnwidth}
  \centering
  \includegraphics[width=\columnwidth]{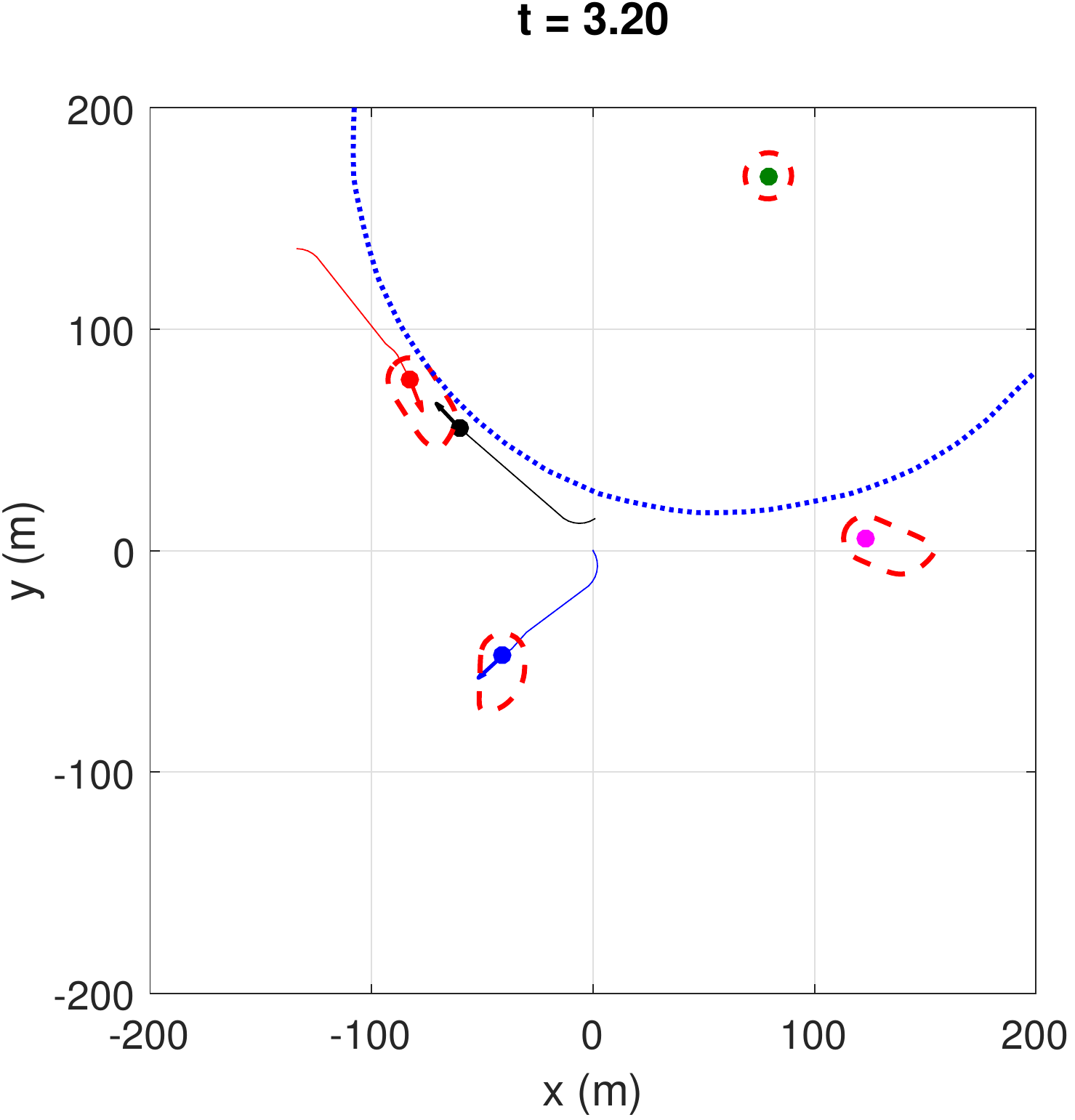}
  \subcaption{}
  \label{fig:worstcase2}
\end{subfigure}
\begin{subfigure}{.5\columnwidth}
  \centering
  \includegraphics[width=\columnwidth]{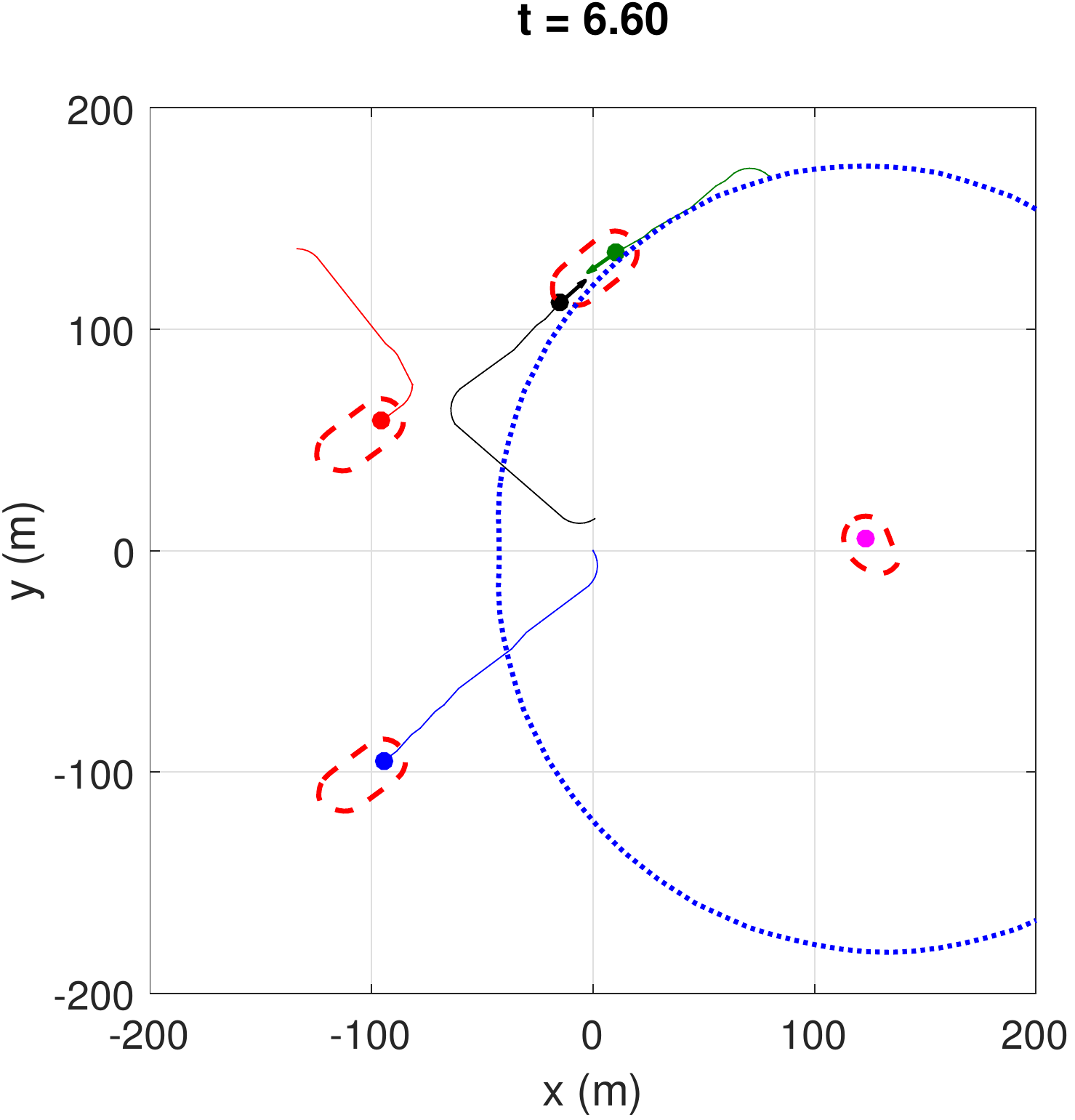}
  \subcaption{}
  \label{fig:worstcase3}
\end{subfigure}%
\begin{subfigure}{.5\columnwidth}
  \centering
  \includegraphics[width=\columnwidth]{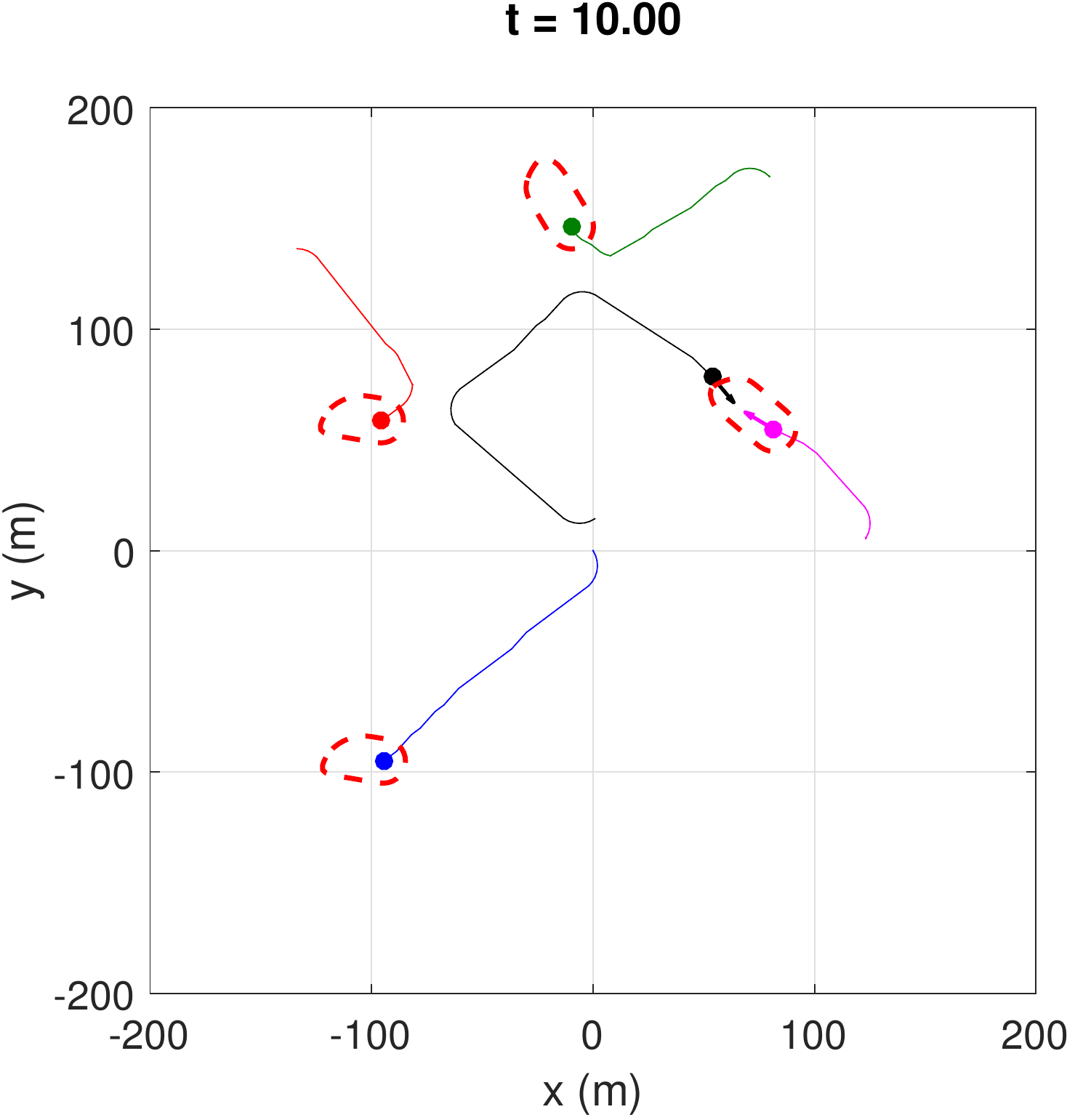}
  \subcaption{}
  \label{fig:worstcase4}
\end{subfigure}
\caption{Illustration of the intruder strategy to force maximum number of vehicles to apply an avoidance maneuver and hence to replan their trajectories. $\veh_{\intr}$ is able to force $\nva=3$ vehicles to apply an avoidance control if the vehicles are applying the \textit{worst control} which takes it closer to the intruder while the intruder is trying to reach its avoid region boundary. }
\label{fig:worstcase}
\end{figure}

Under the proposed algorithm, the intruder will affect the maximum number of vehicles ($\nva$ vehicles), when it appears at the boundary of the avoid region of a vehicle, immediately travels through the buffer region between vehicles and reaches the boundary of the avoid region of another vehicle at $\tsa + \brd$ and then the boundary of the avoid region of another vehicle at $\tsa + 2\brd$ and so on. This strategy will make sure that the intruder forces maximum vehicles to apply an avoidance maneuver during a duration of $\iat$. This is illustrated for a small simulation of 4 vehicles in Figure \ref{fig:worstcase}. In this case at $\tsa = 0$, $\veh_{\intr}$ (Black vehicle) appears at the boundary of the avoid region of $\veh_1$ (Blue vehicle) (see Figure \ref{fig:worstcase1}). Immediately, it travels through the buffer region between $\veh_1$ and $\veh_2$ and at $t = \tsa + \brd = $\SI{3.33}{\s}, reaches the boundary of the avoid region of $\veh_2$ (Red vehicle), as shown in Figure \ref{fig:worstcase1}. The trajectories that $\veh_1$ will follow while applying the avoidance control, and $\veh_2$ and $\veh_{\intr}$ will follow while trying to collide with each other are also shown. Following the same strategy, $\veh_{\intr}$ reaches the boundary of the avoid region of $\veh_3$ (Green vehicle) at $t = \tsa + 2\brd = $\SI{6.67}{\s}, and will just barely reach the boundary of the avoid region of $\veh_4$ (Pink vehicle) at $t = $\SI{10}{\s}. However, it won't be able to force $\veh_4$ to apply an avoidance maneuver as the duration of $\iat$ will be over by then. Thus the avoid start time of the four vehicles are given as $\tsa_1 = $\SI{0}{\s}, $\tsa_2 = $\SI{3.33}{\s}, $\tsa_3 = $\SI{6.67}{\s} and $\tsa_4 = \infty$. The set of vehicles that will need to replan their trajectories after the intruder disappears is given by $\rvs = \{\veh_1, \veh_2, \veh_3\}$. As expected, $|\rvs| \leq 3$.    

\begin{figure}
\centering
\begin{subfigure}{.5\columnwidth}
  \centering
  \includegraphics[width=\columnwidth]{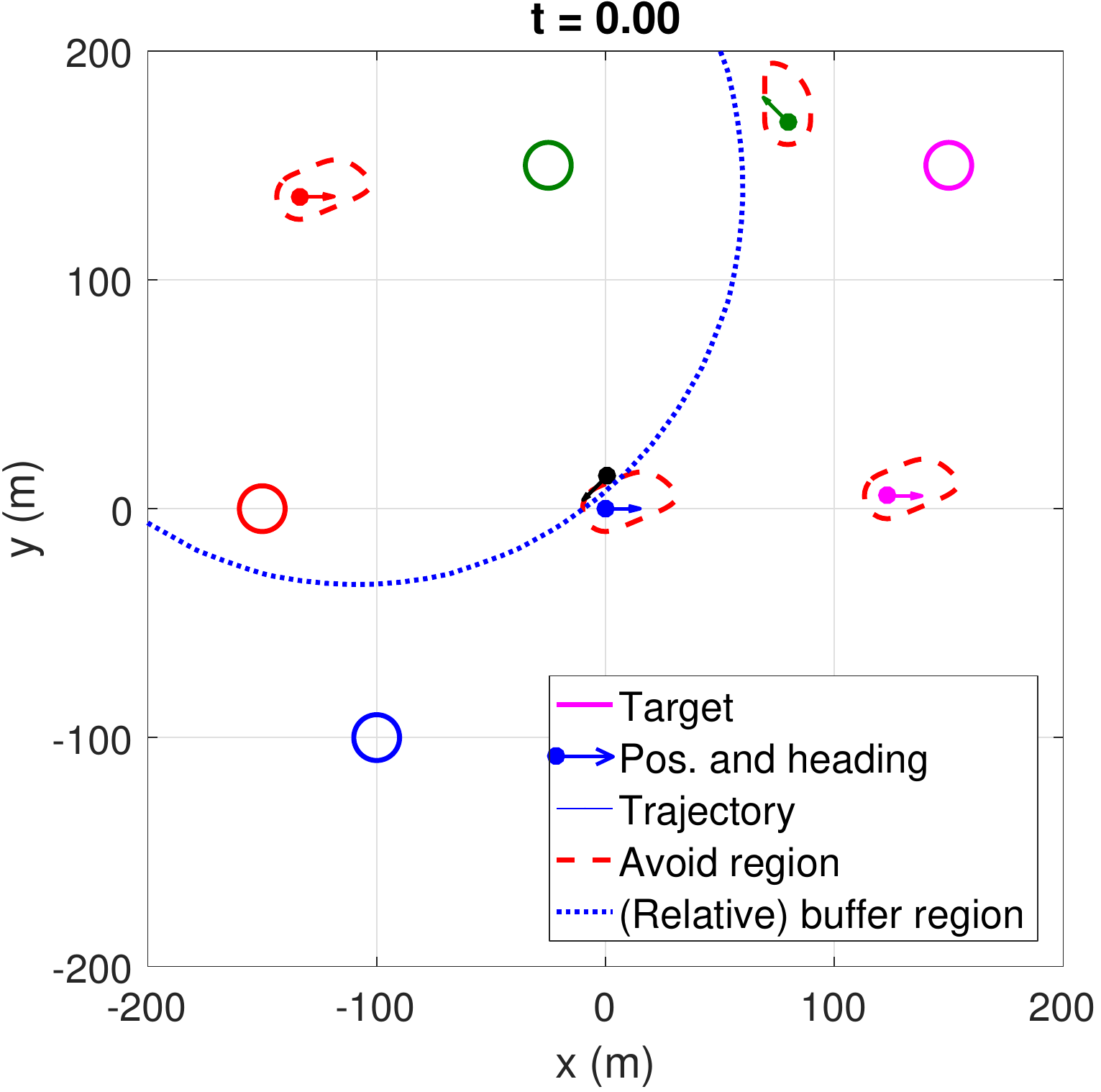}
  \subcaption{}
  \label{fig:normalcase1}
\end{subfigure}%
\begin{subfigure}{.5\columnwidth}
  \centering
  \includegraphics[width=\columnwidth]{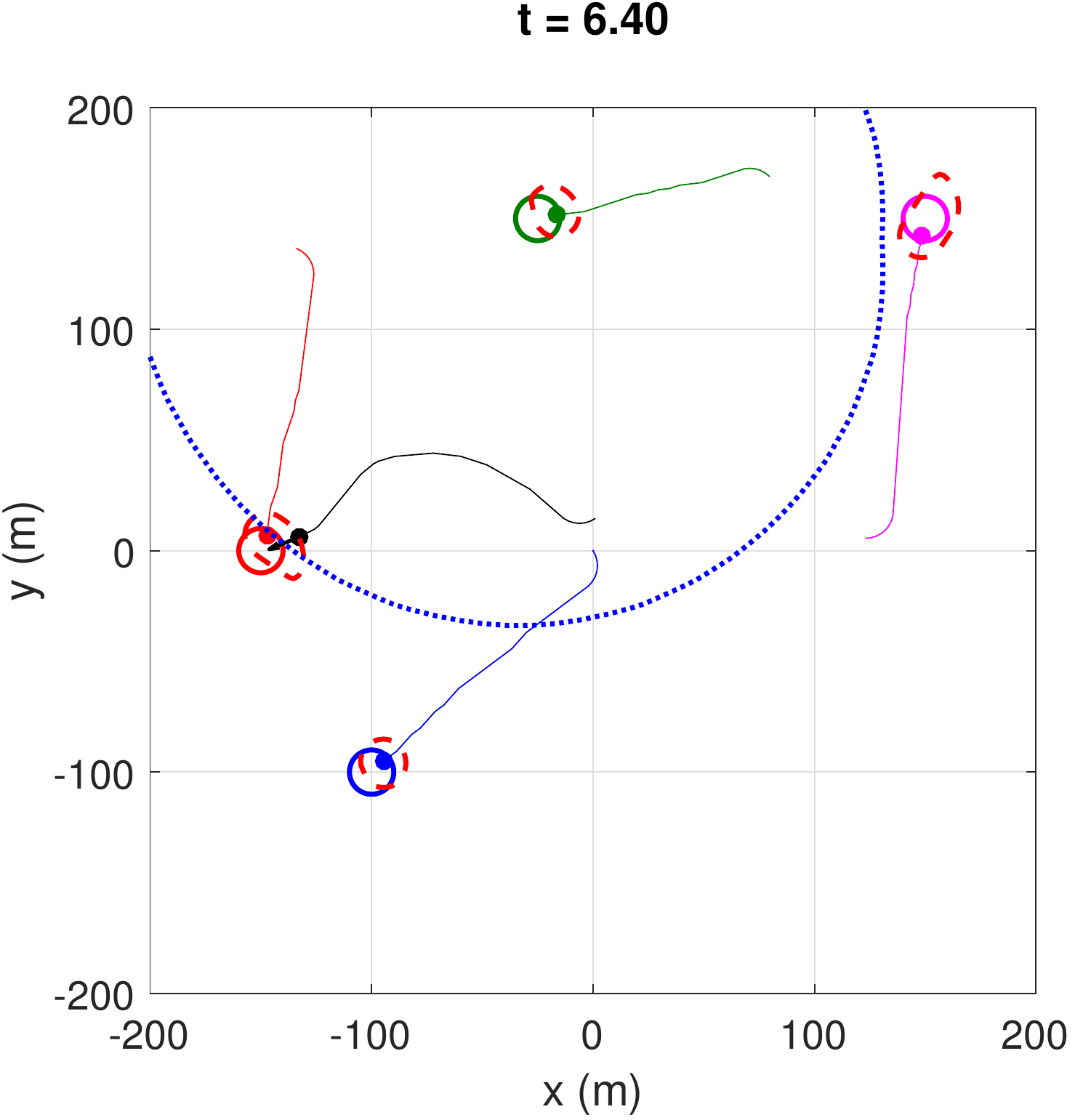}
  \subcaption{}
  \label{fig:normalcase2}
\end{subfigure}
\begin{subfigure}{.5\columnwidth}
  \centering
  \includegraphics[width=\columnwidth]{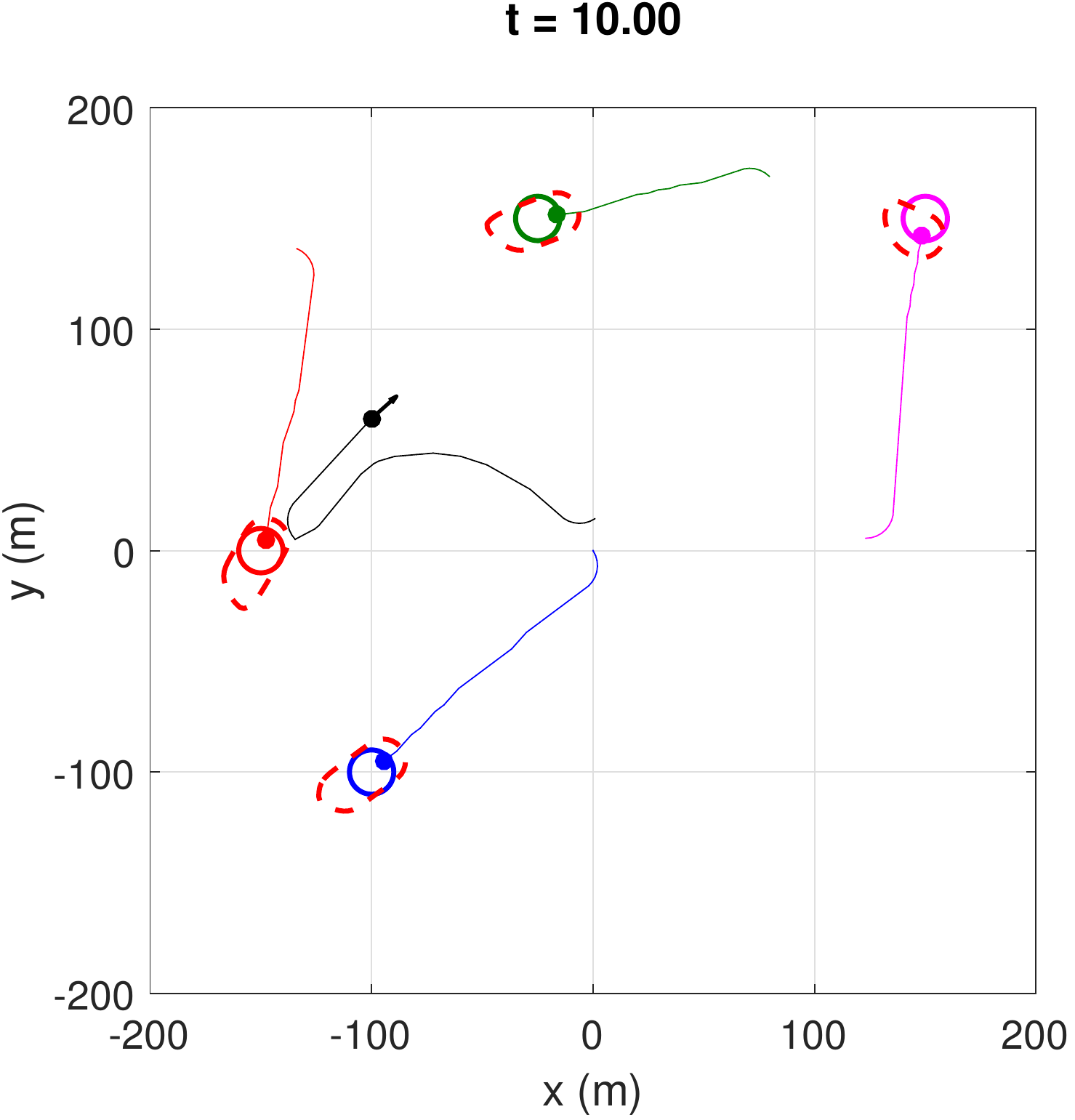}
  \subcaption{}
  \label{fig:normalcase3}
\end{subfigure}%
\caption{Illustration of the intruder strategy to force the maximum number of vehicles to apply an avoidance maneuver and hence to replan their trajectories. Since a vehicle's nominal controller might be different from the worst case controller that is assumed while computing the buffer region, $\veh_{\intr}$ is very likely to be able to force less than $\nva$ vehicles to apply an avoidance maneuver despite its best strategy.}
\label{fig:normalcase}
\end{figure}

The relative buffer region between vehicles is computed under the assumption that both the STP vehicle and the intruder are trying to collide with each other; this is to ensure that the intruder will need at least a duration of $\brd$ to reach the boundary of the avoid region of the next vehicle, irrespective of the control applied by the vehicle. However, a vehicle will be applying the control policy $\ctrl^{\text{PP}}(\cdot)$ unless the intruder forces it to apply an avoidance maneuver, which may not necessarily correspond to the policy that the vehicle will use to \textit{deliberately} collide with the intruder. Therefore, it is very likely that the intruder will need a larger duration to reach the boundary of the avoid region of next vehicle, and hence it will be able to affect less than $\nva$ vehicles even with its best strategy to affect maximum vehicles. This is also evident from Figure \ref{fig:normalcase}. In this case, $\veh_{\intr}$ again appears at the boundary of the avoid region of $\veh_1$ at $t=0$, as shown in Figure \ref{fig:normalcase1}. The respective targets of the vehicles are also shown. Following its best strategy, the intruder immediately moves to travel through the buffer region between $\veh_1$ and $\veh_2$. However, $\veh_2$ now applies the control policy $\ctrl^{\text{PP}}(\cdot)$, i.e. it is trying to reach its target, unless the intruder reaches the boundary of its avoid region, which does not happen until $t= $\SI{6.4}{\s}. Now, intruder again tries to travel through the avoid region of $\veh_2$ and $\veh_3$, but is not able to reach the boundary of the avoid region of $\veh_3$ before it is removed from the system at $t=\iat=$\SI{10}{\s}. Thus, the intruder is able to force only two vehicles to apply an avoidance maneuver. The avoid start time of the four vehicles are given as $\tsa_1 = $\SI{0}{\s}, $\tsa_2 = $\SI{6.4}{\s}, $\tsa_3 = \infty$ and $\tsa_4 = \infty$. The set of vehicles that will need to replan their trajectories is given by $\rvs = \{\veh_1, \veh_2\}$. This conservatism in our method is discussed further in Section \ref{sec:discuss}.

The time for planning and replanning for each vehicle is approximately 15 minutes on a MATLAB implementation on a desktop computer with a Core i7 5820K processor. With a GPU-parallelized CUDA implementation in C++ using two GeForce GTX Titan X graphics processing units, this computation time is reduced to approximately 9 seconds per vehicle. So for $\nva = 3$, replanning would take less than 30 seconds. Reachability computations are highly parallelizable, and with more computational resources, replanning should be possible to do within a fraction of seconds.

\subsection{Discussion \label{sec:discuss}}
The simulations illustrate the effectiveness of reachability in ensuring that the STP vehicles safely reach their respective destinations even in the presence of an intruder. However, they also highlight some of the conservatism in the worst-case reachability analysis. For example, in the proposed algorithm, we assume the worst-case disturbances and intruder behavior while computing the buffer region and the induced obstacles, which results in a large separation between vehicles and hence a lower vehicle density overall, as evident from Figure \ref{fig:trajObsSim}. Similarly, while computing the relative buffer region, we assumed that a vehicle is \textit{deliberately} trying to collide with the intruder so we once again consider the worst-case scenario, even though the vehicle will only be applying the nominal control strategy $\ctrl^{\text{PP}}(\cdot)$, which is usually not be same as the worst-case control strategy. This worst-case analysis is essential to guarantee safety regardless of the actions of STP vehicles, the intruder, and disturbances, given no other information about the intruder's intentions and no model of disturbances except for the bounds. However, the conservatism of our results illustrates the need and the utility of acquiring more information about the intruder and disturbances, and of incorporating knowledge of the nominal strategy $\ctrl^{\text{PP}}(\cdot)$ in future work.

\section{Conclusion and Future Work}
We propose an algorithm to account for an adversarial intruder in sequential trajectory planning. All vehicles are guaranteed to successfully reach their respective destinations without entering each other's danger zones despite the worst-case disturbance and the intruder attack the vehicles could experience. The proposed method ensures that only a fixed number of vehicles need to replan their trajectories once the intruder disappears, irrespective of the total number of vehicles. Moreover, this fixed number is an input to the algorithm and hence can be chosen such that the replanning process is feasible in real-time. The proposed method is illustrated in a fifty-vehicle simulation, set in the urban environment of San Francisco city in California, USA. Future work includes exploring methods that can account for multiple simultaneous intruders and reduce conservatism in the current analysis.

\bibliographystyle{IEEEtran}
\bibliography{IEEEabrv,references}

\section{Appendix}

\subsection{Separation and Buffer Regions - Case 2} \label{sec:case2}
In this section, we consider Case 2: $\tsa_i < \tsa_j, \tsa_i <\infty$. In this case, the intruder forces $\veh_i$, the lower-priority vehicle, to apply an avoidance maneuver before $\veh_j$, the higher-priority vehicle. The analysis in this case is similar to that of Case 1 (Section \ref{sec:case1}). However, there are a few subtle differences, which we point out wherever relevant. We start our analysis with an observation similar to Observation \ref{obs1_case1}:
\begin{observation} \label{obs1_case2}
Without loss of generality, we can assume that $\state_{\intr, i}(\tsa) \in \partial \brs^{\text{A}}_{i}(0, \iat)$. Equivalently, we can assume that $\tsa_i = \tsa$.
\end{observation}
\subsubsection{Separation region} \label{sec:sepRegion_case2}
Similar to Section \ref{sec:sepRegion_case1}, we want to compute the set of all states of the intruder for which $\veh_j$ is forced to apply an avoidance maneuver. Since, $\veh_j$ applies the avoidance maneuver after $\veh_i$ in this case, $\veh_j$ will need to avoid the intruder for a maximum duration of $\trd := \iat - \brd$. This is due to the fact that our design of the buffer region in Section \ref{sec:buffRegion_case2} ensures that it takes the intruder at least a duration of $\brd$ to go from the boundary of the avoid region of $\veh_i$ to that of $\veh_j$. $\sep_j(\tsa_j)$ can thus be obtained as:
\begin{equation} \label{eqn:sepRegion_case2}
\sep_j(\tsa_j) = \boset_j(\tsa_j) + \partial \brs^{\text{A}}_{j}(0, \trd).
\end{equation}

\subsubsection{Buffer Region} \label{sec:buffRegion_case2}
The idea behind the design of buffer region is same as that in Case 1: we want to make sure that $\veh_{\intr}$ spends at least a duration of $\brd$ to go from the boundary of the avoid region of one STP vehicle to the boundary of the avoid region of some other STP vehicle. Mathematically, we want to compute the set of all states $\state_{\intr}$ such that if $\veh_{\intr}$ starts in this set at time $t$, it cannot reach $\sep_j(\cdot)$ before $t_1 = t + \brd$, regardless of the control applied by $\veh_j$ and $\veh_{\intr}$ during interval $[t, t_1]$. Similar to Section \ref{sec:buffRegion_case1}, this set is given by $\brs^{\text{B}}_{j}(0, \brd)$: 
\begin{equation} \label{eqn:buffBRS_case2}
\begin{aligned}
\brs^{\text{B}}_{j}(0, \brd) = & \{y: \exists \ctrl_j(\cdot) \in \cfset_j, \exists \ctrl_\intr(\cdot) \in \cfset_\intr, \exists \dstb_j(\cdot) \in \dfset_j, \\
& \exists \dstb_\intr(\cdot) \in \dfset_\intr, \state_{\intr, j}(\cdot) \text{ satisfies \eqref{eq:reldyn}},\\
& \exists s \in [0, \brd], \state_{\intr, j}(s) \in \brs^{\text{A}}_{j}(\brd, \iat),\\
& \state_{\intr, j}(t) = y\},
\end{aligned}
\end{equation}
where 
\begin{equation}
H^{\text{B}}_{j}(\state_{\intr, j}, \costate) = \min_{\substack{\ctrl_j \in \cset_j, \ctrl_\intr \in \cset_\intr, \\ \dstb_j \in \dset_j, \dstb_\intr \in \dset_\intr}} \costate \cdot f_r(\state_{\intr, j}, \ctrl_j, \ctrl_\intr, \dstb_j, \dstb_\intr)
\end{equation}

In absolute coordinates, we thus have that if the intruder starts outside $\tilde{\buff}_{ji}(t) = \boset_j(t) + \brs^{\text{B}}_{j}(0, \brd)$ at time $t$, then it cannot reach $\sep_j(\cdot)$ before time $t + \brd$. Finally, if we can ensure that the avoid region of $\veh_i$ at time $t$ is outside $\tilde{\buff}_{ji}(t)$, then $\state_{\intr, i}(\tsa_i) \in \partial \brs^{\text{A}}_{i}(0, \iat)$ implies that $\tsa_j - \tsa_i \geq \brd$. Mathematically, if we define the set,  
\begin{equation} \label{eqn:buffRegion_case2}
\buff_{ji}(\tsa) = \boset_j(\tsa) + \brs^{\text{B}}_{j}(0, \brd) + \left(-\brs^{\text{A}}_{i}(0, \iat)\right),
\end{equation} 
then $(\tsa_j - \tsa_i) \geq \brd$ as long as $\state_i(\tsa) \in \left(\buff_{ji}(\tsa)\right)^C$. Thus, if $\state_i(\tsa) \in \left(\buff_{ji}(\tsa)\right)^C$, then the separation requirement \eqref{eqn:sep_cond} is satisfied for Case 2. Further, if $\state_i(\tsa) \in \left(\buff_{ji}(\tsa) \cup \buff_{ij}(\tsa)\right)^C$, then the separation requirement is satisfied regardless of \textit{any} intruder strategy. 

Note that we use $-\brs^{\text{A}}_{i}(0, \iat)$ instead of $\brs^{\text{A}}_{i}(0, \iat)$ in \eqref{eqn:buffRegion_case2} because $\brs^{\text{A}}_{i}(0, \iat)$ is computed using the relative state $\state_{\intr, i}$ and we are interested in finding the ``unsafe" states for $\veh_i$ when the intruder is outside $\tilde{\buff}_{ji}(t)$.
\subsubsection{Obstacle Computation} \label{sec:intruderObs_case2}
We now compute the set of states that $\veh_i$ needs to avoid in order to avoid entering in the danger zone of $\veh_j$ eventually. We consider the following two mutually exclusive and exhaustive cases:
\begin{enumerate}
\item Case A: The intruder affects $\veh_i$, but not $\veh_j$, i.e., $\tsa_i < \infty$ and $\tsa_j = \infty$.
\item Case B: The intruder first affects $\veh_i$ and then $\veh_j$, i.e., $\tsa_i < \tsa_j < \infty$.
\end{enumerate}
For each case, we compute the set of states that $\veh_i$ needs to avoid at time $t$ to avoid entering in $\dz_{ij}$ eventually. We also let $_2^A\ioset_i^j(\cdot)$ and $_2^B\ioset_i^j(\cdot)$ denote the set of obstacles corresponding to Case A and Case B respectively. 
\begin{itemize}[leftmargin=*] 
\item \label{sec:intruderObs_case2_caseA} Case A: In this case, we need to ensure that $\veh_i$ does not collide with $\veh_j$ while it is avoiding the intruder. Since $\veh_j$ is not avoiding the intruder in this particular case, the set of possible states of $\veh_j$ at time $t$ is given by $\boset_j(t)$. To compute $_2^A\ioset_i^j(\cdot)$, we begin with the following observation: 
\begin{observation} \label{obs1_case2_caseA}
By Observation \ref{obs1_case1Obs}, it is sufficient to consider the scenarios where $\tsa = \tsa_i \in [t-\iat, t]$. Since $\veh_i$ is forced to apply an avoidance maneuver for the time interval $[\tsa_i, \tsa_i+\iat]$, it needs to be ensured that $\veh_i$ avoids all states at time $t$ that can lead to a collision with $\veh_j$ during the interval $[t, \tsa_i+\iat]$ for some avoidance control. Therefore, it is sufficient to consider the scenario $\tsa_i = t$, as it will maximize the avoidance duration $[\tsa_i, \tsa_i+\iat]$ for the obstacle computation at time $t$.  
\end{observation}

Mathematically, $\veh_i$ needs to avoid all states at time $t$ that can reach $\mathcal{K}^{\text{A2}}(\tau)$ for some control action of $\veh_i$ during time duration $[t, \tau]$. $\mathcal{K}^{\text{A2}}(\tau)$ here is given by:
\begin{equation}
\begin{aligned}
\mathcal{K}^{\text{A2}}(\tau) = & \tilde{\boset}_j(\tau),\\
\tilde{\boset}_j(s) = & \{\state_j: \exists (y, h) \in \boset_j(s), \|\pos_j - y\|_2 \le \rc \}.
\end{aligned}
\end{equation}
$\tilde{\boset}_j(s)$ represent the set of all states that are in potential collision with $\veh_j$ at time $s$. Note that since the intruder is present in the system for a maximum duration of $\iat$ and since $\tsa_i = t$ (by Observation \ref{obs1_case2_caseA}), we have that $\tau \in [t, t+\iat]$. 

Avoiding $\mathcal{K}^{\text{A2}}(\cdot)$ will ensure that $\veh_i$ and $\veh_j$ will not enter into each other's danger zones regardless of the avoidance maneuver applied by $\veh_i$. The set of states that $\veh_i$ needs to avoid at time $t$ is thus given by the following BRS: 
\begin{equation} \label{eq:ObsBRS_case2_caseA}
\begin{aligned}
\brs^{\text{A2}}_{i}(t, t+\iat) = & \{y: \exists \ctrl_i(\cdot) \in \cfset_i, \exists \dstb_i(\cdot) \in \dfset_i, \\
& \state_i(\cdot) \text{ satisfies \eqref{eq:dyn}}, \state_i(t) = y, \\
& \exists s \in [t, t+\iat], \state_i(s) \in \mathcal{K}^{\text{A2}}(s) \}.
\end{aligned}
\end{equation}
The Hamiltonian $\ham^{\text{A2}}_{i}$ to compute $\brs^{\text{A2}}_{i}(t, t+\iat)$ is given by:
\begin{equation} \label{eqn:BRS_obsham_case2_caseA}
\ham^{\text{A2}}_{i}(\state_i, \costate) = \min_{\ctrl_i \in \cset_i, \dstb_i \in \dset_i} \costate \cdot f_i (\state_i, \ctrl_i, \dstb_i).
\end{equation}
$\brs^{\text{A2}}_{i}(t, t+\iat)$ represents the set of all states of $\veh_i$ at time $t$ from which it is possible for $\veh_i$ to reach $\mathcal{K}^{\text{A2}}(\tau)$ for some $\tau \geq t$. Thus, the induced obstacle in this case is given as
\begin{equation} \label{eq:intruderObs_case2_caseA}
_2^A\ioset_i^j(t) = \brs^{\text{A2}}_{i}(t, t+\iat).
\end{equation}

\item \label{sec:intruderObs_case2_caseB} Case B: In this case, the intruder first affects $\veh_i$ and then $\veh_j$. Recall that $\veh_i$ and $\veh_j$ apply their first avoidance maneuver at $\tsa_i$ and $\tsa_j$ respectively. Since the intruder appears for a maximum duration of $\iat$ and $\tsa_i = \tsa$, from the perspective  of $\veh_i$, $\veh_j$ can apply any control during the duration $[\tsa_j, \tsa_i + \iat]$ and hence can be anywhere in the set $\frs_{j}^{\mathcal{O}}(\tsa_j, \tau)$ at $\tau \in [\tsa_j, \tsa_i + \iat]$, where $\frs_{j}^{\mathcal{O}}$ denotes the FRS of base obstacle $\boset_j(\tsa_j)$ computed forward for a duration of $(\tsa_i + \iat - \tsa_j)$. $\veh_i$ thus needs to make sure that it avoids all states at time $t$ that can reach $\frs_{j}^{\mathcal{O}}(\tsa_j, \tau)$, regardless of the control applied by $\veh_i$ during $[t, \tau]$. We now make the following key observation:
\begin{observation} \label{obs1_case2_caseB}
Observation \ref{obs1_case1_caseA} implies that $\frs_{j}^{\mathcal{O}}(\tau_2, \tau) \subseteq \frs_{j}^{\mathcal{O}}(\tau_1, \tau)$ if $\tau > \tau_2 > \tau_1$. Therefore, the biggest obstacle, $\frs_{j}^{\mathcal{O}}(\tsa_j, \tau)$, is induced by $\veh_j$ at $\tau$ if $\tsa_j$ is as early as possible. Hence, it is sufficient for $\veh_i$ to avoid this biggest obstacle to ensure collision avoidance with $\veh_j$ at time $\tau$. Given the separation and buffer regions between $\veh_i$ and $\veh_j$, we must have $\tsa_j - \tsa_i \geq \brd$. Hence, the biggest obstacle is induced by $\veh_j$ when $\tsa_j = \tsa_i + \brd$. 
\end{observation}
Intuitively, Observation \ref{obs1_case2_caseB} implies that the biggest obstacle is induced by $\veh_j$ when intruder forces $\veh_i$ to apply the avoidance maneuver and \textit{immediately} begins traveling through the buffer region between two vehicles to force $\veh_j$ to apply an avoidance maneuver after a duration of $\brd$. Therefore, $\veh_i$ needs to avoid $\mathcal{K}^{\text{B2}}(\tau)$ at time $\tau > t$, where 
\begin{equation} \label{eqn:obs_case2_caseB_help1}
\mathcal{K}^{\text{B2}}(\tau) =  \bigcup_{\tsa_i \in [t-\iat, t], \tau \leq \tsa_i + \iat} \frs_{j}^{\mathcal{O}}(\tsa_i + \brd, \tau), \tau>t,
\end{equation}
where we have substituted $\tsa_j = \tsa_i + \brd$. In \eqref{eqn:obs_case2_caseB_help1}, $\tsa = \tsa_i \in [t-\iat, t]$ due to Observation \ref{obs1_case1Obs} and $\tau \leq \tsa_i + \iat$ because the intruder can appear for a maximum duration of $\iat$. Equation \eqref{eqn:obs_case2_caseB_help1} can be equivalently written as:  
\begin{equation} \label{eqn:obs_case2_caseB_help2}
\begin{aligned}
\mathcal{K}^{\text{B2}}(\tau) = & \bigcup_{\tsa_i \in [\tau-\iat, t]} \frs_{j}^{\mathcal{O}}(\tsa_i + \brd, \tau), t < \tau \leq t + \iat\\
\mathcal{K}^{\text{B2}}(\tau) = & \frs_{j}^{\mathcal{O}}(\tau-\iat+\brd, \tau), t < \tau \leq t + \iat,
\end{aligned}
\end{equation}
where the second equality holds because of Observation \ref{obs1_case1_caseA}. The set of states that $\veh_i$ needs to avoid at time $t$ is thus given by the following BRS:  
\begin{equation} \label{eq:ObsBRS_case2_caseB}
\begin{aligned}
\brs^{\text{B2}}_{i}(t, t+\iat) = & \{y: \exists \ctrl_i(\cdot) \in \cfset_i, \exists \dstb_i(\cdot) \in \dfset_i, \\
& \state_i(\cdot) \text{ satisfies \eqref{eq:dyn}}, \state_i(t) = y, \\
& \exists s \in [t+\brd, t+\iat], \state_i(s) \in \tilde{\mathcal{K}}^{\text{B2}}(s) \},\\
\tilde{\mathcal{K}}^{\text{B2}}(s) = & \{\state_i: \exists (y, h) \in \mathcal{K}^{\text{B2}}(s), \|\pos_i - y\|_2 \le \rc \}.
\end{aligned}
\end{equation}
The Hamiltonian $\ham^{\text{B2}}_{i}$ to compute $\brs^{\text{B2}}_{i}(t, t+\iat)$ is given by:
\begin{equation} \label{eqn:BRS_obsham_case2_caseB}
\ham^{\text{B2}}_{i}(\state_i, \costate) = \min_{\ctrl_i \in \cset_i, \dstb_i \in \dset_i} \costate \cdot f_i (\state_i, \ctrl_i, \dstb_i).
\end{equation}
Finally, the induced obstacle in this case is given as
\begin{equation} \label{eq:intruderObs_case2_caseB}
_2^B\ioset_i^j(t) = \brs^{\text{B2}}_{i}(t, t+\iat).
\end{equation}
\end{itemize}

\end{document}